\pdfoutput=1
\documentclass{sig-alternate}
\usepackage{color}
\usepackage{caption}
\usepackage{subcaption}
\usepackage{booktabs}
\newcommand{\ra}[1]{\renewcommand{\arraystretch}{#1}}
\usepackage{url}
\usepackage{verbatim} 
\usepackage{multirow}
\usepackage{algorithm}
\usepackage[noend]{algorithmic}
\usepackage{xspace}
\usepackage{graphicx}
\usepackage{epsfig}
\usepackage{amsmath}
\usepackage{amssymb}
\usepackage{times}
\usepackage{dsfont}
\usepackage{paralist}
\usepackage{balance}

\newcommand{\hide}[1]{}
\newcommand{\xhdr}[1]{\vspace{1mm}\noindent{{\bf #1.}}}

\newcommand{\denselist}{ \itemsep -2pt\topsep-10pt\partopsep-10pt }

\graphicspath{{./FIG/}}

\permission{}
\conferenceinfo{}{}
\copyrightetc{}
\crdata{}

\begin{document}

\title{Can Cascades be Predicted?}

\numberofauthors{5} 

\author{
\alignauthor
Justin Cheng\\
      \affaddr{Stanford University}\\
      \email{jcccf@cs.stanford.edu}
\alignauthor
Lada A. Adamic\\
      \affaddr{Facebook}\\
      \email{ladamic@fb.com}
\alignauthor
P. Alex Dow\\
      \affaddr{Facebook}\\
      \email{adow@fb.com}
\and
\alignauthor Jon Kleinberg\\
      \affaddr{Cornell University}\\
      \email{kleinber@cs.cornell.edu}
\alignauthor Jure Leskovec\\
      \affaddr{Stanford University}\\
      \email{jure@cs.stanford.edu}
}

\maketitle

\begin{abstract}

On many social networking web sites such as Facebook and Twitter,
resharing or reposting functionality allows users to share others'
content with their own friends or followers. As content is
reshared from user to user, large cascades of reshares can form.
While a growing body of research has focused on analyzing and
characterizing such cascades, a recent, parallel line of work has 
argued that the future trajectory of a cascade may be inherently
unpredictable.
In this work, we develop a framework for addressing cascade prediction
problems.
On a large sample of photo reshare cascades on Facebook, we find strong performance in predicting whether a cascade will continue to grow in the future.
We find that the relative growth of a cascade becomes more predictable as we observe more of its reshares, that temporal and structural features are key predictors of cascade
size, and that initially, breadth, rather than depth in a cascade is a better
indicator of larger cascades.
This prediction performance is robust in the sense that multiple distinct classes of features all achieve similar performance.
We also discover that temporal features are predictive of a cascade's eventual shape.
Observing independent cascades of the same content, we find that while
these cascades differ greatly in size, we are still able to predict which ends up the largest.

\end{abstract}


\vspace{1mm}
\noindent {\bf Categories and Subject Descriptors:} H.2.8 {\bf
[Database Management]}: Database applications---{\it Data mining}

\noindent {\bf General Terms:} Experimentation, Measurement.


\section{Introduction}
\label{sec:intro}

The sharing of content through social networks has become an
important mechanism by which people discover and consume information online.
In certain instances, a photo, link, or other piece of information
may get {\em reshared} multiple times: a user shares the content with her
set of friends, several of these friends share it with their respective
sets of friends, and a {\em cascade} of resharing can develop,
potentially reaching a large number of people.  
Such cascades have been identified in settings including
blogging \cite{adar-blogspace,gruhl-blogspace,leskovec-blogspace-sdm07}, 
e-mail \cite{golub-chain-letter,liben-nowell-pnas08}, 
product recommendation \cite{leskovec-ec06j},
and social sites such as Facebook and Twitter
\cite{dow2013anatomy,kumar-conversations}.
A growing body of research has focused on characterizing 
cascades in these domains, 
including their structural properties and their content.

In parallel to these investigations, there has been a recent line of 
work adding notes of caution to the study of cascades.
These cautionary notes fall into two main genres: first, that
large cascades are rare \cite{goel2012structure};
and second, that the eventual scope
of a cascade may be an inherently unpredictable property
\cite{salganik-music,watts-everything-book}.
The first concern --- that large cascades are rare --- is a widespread
property that has been observed quantitatively in many systems where
information is shared.
The second concern is arguably more striking, but also much
harder to verify quantitatively: 
to what extent is the future trajectory of a cascade predictable;
and which features, if any, are most useful for this prediction task?

Part of the challenge in approaching this prediction question
is that the most direct ways of formulating it do not fully address
the two concerns above.
Specifically, if we are presented with a short initial portion of
a cascade and asked to estimate its final size, then we are faced
with a pathological prediction task, since almost all cascades are small.
Alternately, if we radically overrepresent large cascades in our sample,
we end up studying an artificial setting that does not resemble
how cascades are encountered in practice.
A set of recent initial studies have undertaken versions of cascade
prediction despite these difficulties
\cite{kupavski2012cikm,ma2013predicting,petrovic2011rt,szabo2010predicting},
but to some extent they are inherent in these problem formulations.

These challenges reinforce the fact that finding a robust way to formulate
the problem of cascade prediction remains an open problem.
And because it is open, we are missing a way to obtain
a deeper, more fundamental understanding of the predictability of cascades.
How should we set up the question
so that it becomes possible to address these issues directly, 
and engage more deeply with arguments
about whether cascades might, in the end, be inherently unpredictable?

\paragraph*{\bf The present work: Cascade growth prediction}
In this paper, we propose a new approach to the prediction
of cascades, and show that it leads to strong and robust prediction results.
We are motivated by a view of cascades as complex dynamic objects
that pass through successive stages as they grow.
Rather than thinking of a cascade as something whose final endpoint
should be predicted from its initial conditions, we think of it as
something that should be {\em tracked} over time, via a sequence of
prediction problems in which we are constantly seeking to estimate
the cascade's next stage from its current one.

What would it mean to predict the ``next stage'' of a cascade?
If we think about all cascades that reach size $k$, there is a distribution
of eventual sizes that these cascades will reach. Then the distribution
of cascade sizes has a median value $f(k) \geq k$.  This number $f(k)$ is thus the
``typical'' final size for cascades that reached size at least $k$.
Hence, the most basic way to ask about a cascade's next stage of
growth, given that it currently has size $k$, is to ask whether it
reaches size $f(k)$.

We therefore propose the following {\em cascade growth prediction problem}:
given a cascade that currently has size $k$, predict whether it
grow beyond the median size $f(k)$. 
(As we show later, the prediction problem is equivalent to asking: given a cascade of size $k$, will the cascade double its size and reach at least $2k$ nodes?)
This implicitly defines a family of prediction problems, one for each $k$.
We can thus ask how cascade predictability behaves as we sweep over
larger and larger values of $k$.
(There are natural variants and generalizations in which we ask about
reaching target sizes other than the median $f(k)$.)
This problem formulation has a number of strong advantages over standard
ways of trying to define cascade prediction. First, it leads to
a prediction problem in which the classes are balanced, rather than
highly unbalanced. Second, it allows us to ask for the first time how
the predictability of a cascade varies over the range of its growth
from small to large. Finally, it more closely approximates the 
real tasks that need to be solved in applications for
managing viral content, where
many evolving cascades are being monitored, and the question is which
are likely to grow significantly as time moves forward.

For studying cascade growth prediction, it is important to work with a
system in which the sharing and resharing of information is widespread,
the complete trajectories of many cascades---both large and small---are observable, 
and the same piece of content shared separately by many people,
so that we can begin to control for variation in content.
For this purpose, we use a month of complete photo-resharing data from Facebook, 
which provides a rich ecosystem of shared content exhibiting all of these properties.

In this setting, we focus on several categories of questions:
\begin{itemize}
\denselist
\item[(i)] How high an accuracy can we achieve for cascade growth prediction?
If we cannot improve on baseline guessing, then
this would be evidence for the inherent unpredictably of cascades.
But if we can significantly improve on this baseline, then there is a basis for non-trivial prediction.
In the latter case, it also becomes important to understand the
features that make prediction possible.
\item[(ii)] Is growth prediction more tractable on small cascades or 
large ones?  In other words, does the future behavior of a cascade become
more or less predictable as the cascade unfolds?  
\item[(iii)] Beyond just the growth of a cascade, can we predict
its ``shape'' --- that is, its network structure?
\end{itemize}

\paragraph*{\bf Summary of results}
Given the challenges in predicting cascades, 
we find surprisingly strong performance for the growth prediction problem.
Moreover, the performance is robust in the sense that multiple
distinct classes of features, including those based on time, graph structure, and properties of the individuals resharing, can achieve accuracies well above the baseline.
Cascades whose initial reshares come quickly are more likely
to grow significantly; and from a structural point of view, 
breadth rather than depth in the resharing tree is a better
predictor of significant growth.

We investigate the performance of growth prediction as a function
of the size of the cascade so far --- when we want to predict the
growth of a cascade of size $k$, how does our accuracy depend on $k$?
%
It is not a priori clear whether accuracy should increase or
decrease as a function of $k$, since for any value of $k$ the
challenge is to determine what the cascade will do in the future. Seeing
more of the cascade (larger $k$) does not make the problem easier,
as it also involves predicting ``farther'' into the future 
(i.e., whether the cascade will reach size at least $2k$).
We find that accuracy increases with $k$, so that it is possible
to achieve better performance on large cascades than small ones.
The features that are most significant for prediction change with $k$
as well, with properties of the content and the original author
becoming less important, and temporal features remaining relatively stable.

We also consider a related question: how much of a cascade do we need
to see in order to obtain good performance?
Specifically, suppose we want to predict the growth of a cascade of size
at least $R$, but we are only able to see the first $k < R$ nodes
in the cascade.  How does prediction performance depend on $k$, and 
in particular, is there a ``sweet spot'' where a relatively small value
of $k$ gives most of the performance benefits?  
We find in fact that there is no sweet spot: performance essentially
climbs linearly in $k$, all the way up to $k = R$.
Perhaps surprisingly, more information about the cascade continues
to be useful even up to the full snapshot of size $R$.

In addition to growth, we also study how well we
can predict the eventual ``shape'' of the cascade,
using metrics for evaluating tree structures as a numerical
measure of the shape.
We obtain performance significantly above baseline for this task as
well; and perhaps surprisingly, multiple classes of features including
temporal ones perform well for this task, despite the fact that
the quantity being predicted is a purely structural one.

One of the compelling arguments that originally brought the issue of
inherent unpredictability onto the research agenda was a 
striking experiment by Salganik, Dodds, and Watts, in which they
showed that the same piece of content could achieve very different
levels of popularity in separate independent settings
\cite{salganik-music}.
Given the richness of our data, we can study a version of 
this issue here in which we can control
for the content being shared by analyzing many cascades all arising
from the sharing of the same photo.
As in the experiment of Salganik et al., we find that independent resharings of the same photo can generate cascades of very different sizes.
But we also show that this observation can be compatible with prediction:
after observing small initial portions of these distinct cascades
for the same photo, we are able to predict with strong performance
which of the cascades will end up being the largest.
In other words, our data shows wide variation in cascades for the same
content, but also predictability despite this variation.

Overall, our goal is to set up a framework in which prediction
questions for cascades can be carefully analyzed, and our results
indicate that there is in fact a rich set of questions here,
pointing to important distinctions between different types of
features characterizing cascades, and between the
essential properties of large and small cascades.

\section{Related Work}
\label{sec:related}
Many papers have analyzed and cataloged properties of empirically observed information cascades, while others have considered theoretical models of cascade formation in networks.
Most relevant to our work are those which focus on predicting the future popularity of a given piece of content.
These studies have proposed rich sets of features for prediction, which we discuss later in Section~\ref{sec:features}.

Much prior work aims to predict the \emph{volume} of \emph{aggregate} activity --- the total number of up-votes on Digg stories~\cite{szabo2010predicting}, total hourly volume of news phrases~\cite{yang10lim}, or total daily hashtag use~\cite{ma2013predicting}. At the other end of the spectrum, research has focused on \emph{individual} user-level prediction tasks: whether a user will retweet a specific tweet \cite{petrovic2011rt} or share a specific URL \cite{galuba2010outtweeting}.
Rather than attempt to predict aggregate popularity or individual behavior in the next time step, we instead look at whether an information cascade grows over the median size (or doubles in size, as we later show).

Research on communities defined by user interests \cite{backstrom-kdd06} or hashtag content \cite{romero-icwsm13} has also looked at a notion of growth, predicting whether a group will increase in size by a given amount. Nevertheless, these focused on groups of already non-trivial size, and their growth predicted without an explicit internal cascade topology, and without tracking predictability over different size classes.

Several papers focus on predictions after having observed a cascade for a given fixed time frame~\cite{kupavski2012cikm,ma2013predicting,tsur2012s}.
In contrast, rather than studying specific time slices, we continuously observe the cascade over its entire lifetime and attempt to understand how predictive performance varies as the cascade develops. Moreover, our methodology does not penalize slowly but persistently growing cascades. Thus, we predict the size and the structure after having observed a certain number of initial reshares.

Many studies consider the cascade prediction task as a regression problem~\cite{bakshy2009social,kupavski2012cikm,szabo2010predicting,tsur2012s} or a binary classification problem with large bucket sizes~\cite{Hong2011www,jenders13analyzing,kupavski2012cikm}. The danger with these approaches is that they are biased towards studying extremely large but also extremely rare cascades, bypassing the whole issue about the general predictability of cascades.
For example, research has specifically focused on content and users that create extremely large cascades, such as popular hashtags \cite{hoang2012virality,yang2010predicting} and very popular users \cite{dow2013anatomy,guerini2013exploring}, which has led to criticism that cascades may only be predictable after they have already grown large~\cite{watts-everything-book}.
While it is useful to understand the dynamics of extremely popular content, such content is also very rare. Thus, we rather seek to understand predictability along cascade's entire lifetime. We consider cascades that have as few as five reshares, and introduce a classification task which is not skewed towards very large cascades. 



\section{Predicting Cascade Growth}
\label{sec:prediction}
To examine the cascade growth prediction problem, we first define and motivate our experimental setup and the feature sets used, then report our prediction results with respect to different $k$.

\subsection{Experimental setup}

\xhdr{Mechanics of information passing on Facebook}
We focus on content consisting of posts the author has
designated as public, meaning that anyone on Facebook is eligible to view it,
and we further restrict our attention to content in the form of
photos, which comprise the majority of reshare cascades on
Facebook~\cite{dow2013anatomy}.
Such posts are then distributed by Facebook's News Feed, typically
at first to users who are either friends of 
the poster or who subscribe to their content, e.g. as followers.
Each post
is accompanied by a ``share'' link that allows friends and followers
to ``reshare'' the post with her own friends and followers, thus
expanding the set of users exposed to the content. This explicit
sharing mechanism creates information cascades, starting with the root
node (user or page) that originally created the content, and
consisting of all subsequent reshares of that content. 

\begin{figure}[t]
	\centering
	\includegraphics[width=\linewidth]{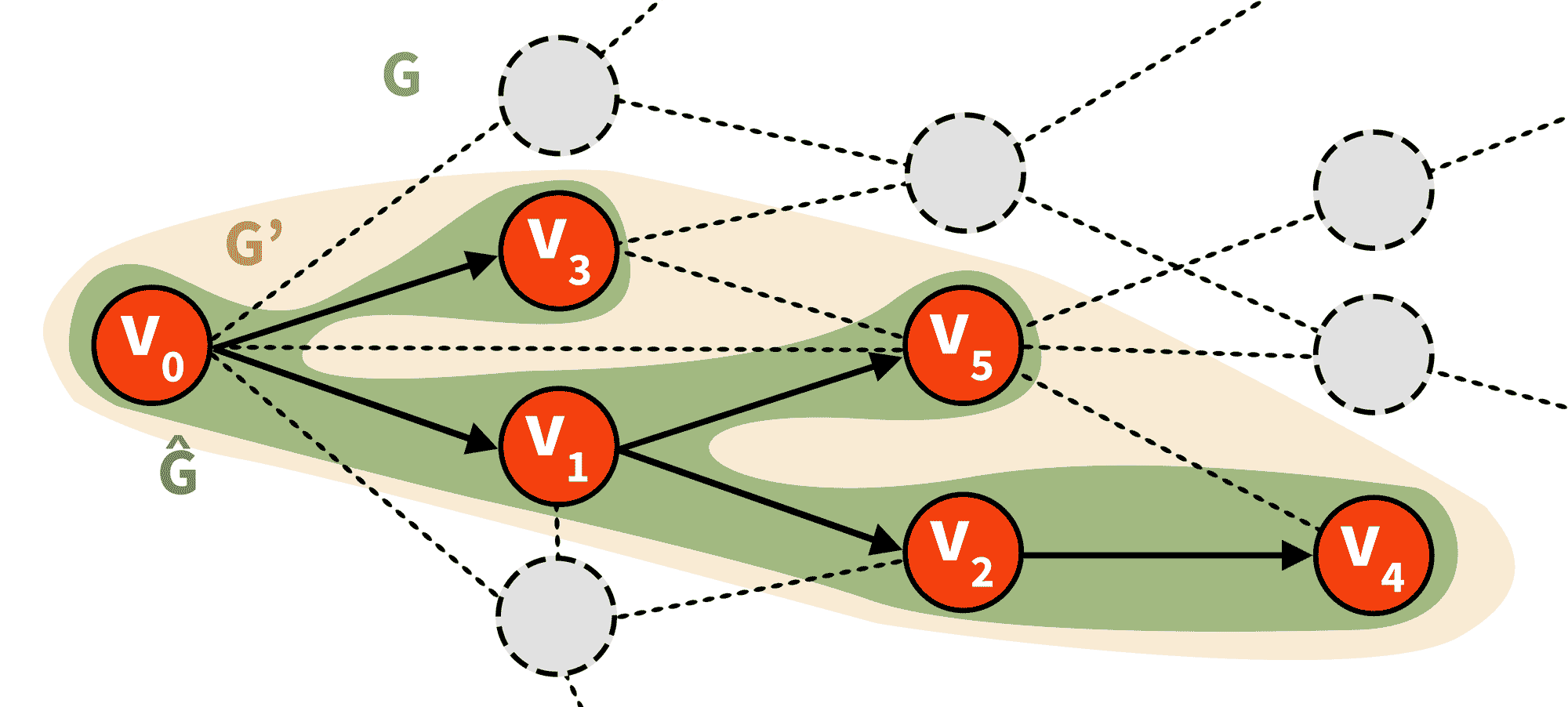}
	\vspace{-5mm}
\caption{An information cascade represented by solid edges on a graph $G$, starting at $v_0$ ($\hat G$). Dashed lines indicate friendship edges; the edges between resharers make up the friend subgraph $G'$.}
\label{fig:cascade}
\vspace{-3mm}
\end{figure}

Figure~\ref{fig:cascade} illustrates the process with an example: a
node $v_0$ posts a public photo, seen by $v_0$'s friends 
and followers in their News Feeds.
Friends $v_1$ and $v_3$ then share the photo with their own friends.
This way the photo propagates over the edges of the Facebook network
and creates an information cascade. We represent the cascade graph as
$\hat G$, and the induced subgraph of all photo sharers, including all
friendship or follow links between them as $G'$. Notice that some
users (ex. $v_5$) are exposed via multiple sources ($v_0, v_1, v_3,
v_4$).

An important issue for our understanding of reshare cascades
is the following distinction: content can be produced by
{\em users} --- individual Facebook accounts whose primary 
audience consists of friends and any subscribers the individual has ---
and it can also be produced by {\em pages}, which correspond to 
the Facebook accounts of companies, brands, celebrities, and other
highly visible public entities.
In the common parlance around cascades, reshared content originally
produced by a user is often informally viewed
as more ``organic,'' developing a following in
a more bottom-up way.
In contrast, reshared content from pages is seen as more top-down,
and generally broadcast via News Feed to a larger set of initial followers.
A natural question, and a theme that will run through several analyses
in the paper, is to understand if these distinctions carry over
to the properties we study here: do user-initiated cascades differ in
their predictability and their underlying structure from 
page-initiated cascades?

\xhdr{Dataset description}
We sampled our anonymized dataset from photos uploaded to Facebook in June 2013 and observed any reshares occurring within 28 days of initial upload. The dataset only includes photos posted publicly (viewable by anyone), and not deleted during the observation period. Further, we exclude photos with fewer than five reshares as is required by the prediction tasks described below. We constructed diffusion trees first by taking the explicit cascade, e.g. C clicking ``share" on B's ``share" of A's photo forms the cascade $A \rightarrow B \rightarrow C$. However, it is possible that user C clicked on user B's share, and then directly reshared from A. Since we want to know how the information actually flowed in the network, we reconstruct the path $A \rightarrow B \rightarrow C$ based on click, impression, and friend/follower data~\cite{dow2013anatomy}.



\begin{figure}[t]
	\centering
	\includegraphics[width=\linewidth]{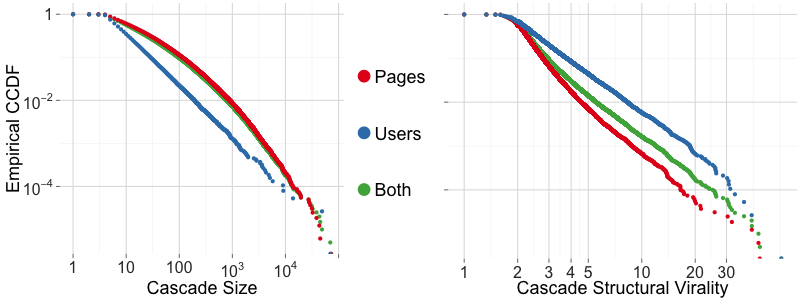}
	\vspace{-4mm}
	\caption{The complementary cumulative distribution (CCDF) of cascade size (left) and structural virality measured by using the Wiener index (right).}
	\label{fig:data}
	\vspace{-5mm}
\end{figure}

Figure \ref{fig:data} begins to show how photos uploaded by pages 
generate cascades that differ from those uploaded by users.
In our dataset, 81\% of cascades are initiated by pages.
Figure~\ref{fig:data} shows the cascade size distribution for pages,
users, and the two combined. Page cascades are typically larger than
user cascades, e.g., 11\% of page cascades reach at least 100
reshares, while only 2\% of user cascades do, though both follow heavy
tailed distributions. Fitting power-law curves to their tails, we observe 
power-law exponents of $\alpha$ equal to 2.2, 2.1, and 2.1 for user,
page, and both, respectively ($x_{\min} =$
10, 2000, 2000).

\begin{figure}
	\centering
	\begin{subfigure}[b]{0.31\linewidth}
	\centering
	\includegraphics[width=0.7\linewidth]{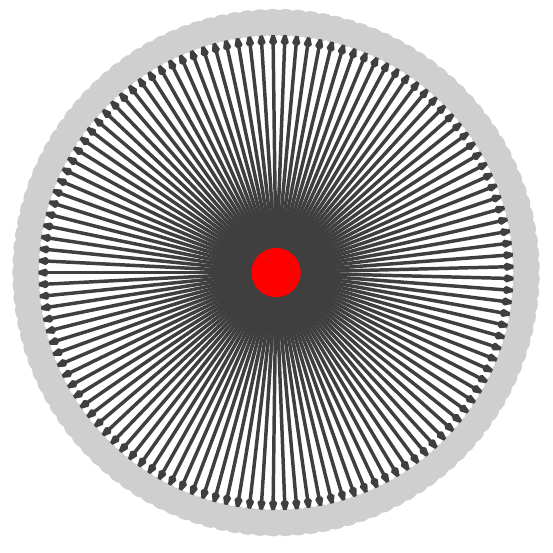}
	\caption{$d=1.98$}
	\end{subfigure}%
	~ 
	\begin{subfigure}[b]{0.31\linewidth}
	\centering
	\includegraphics[width=1\linewidth]{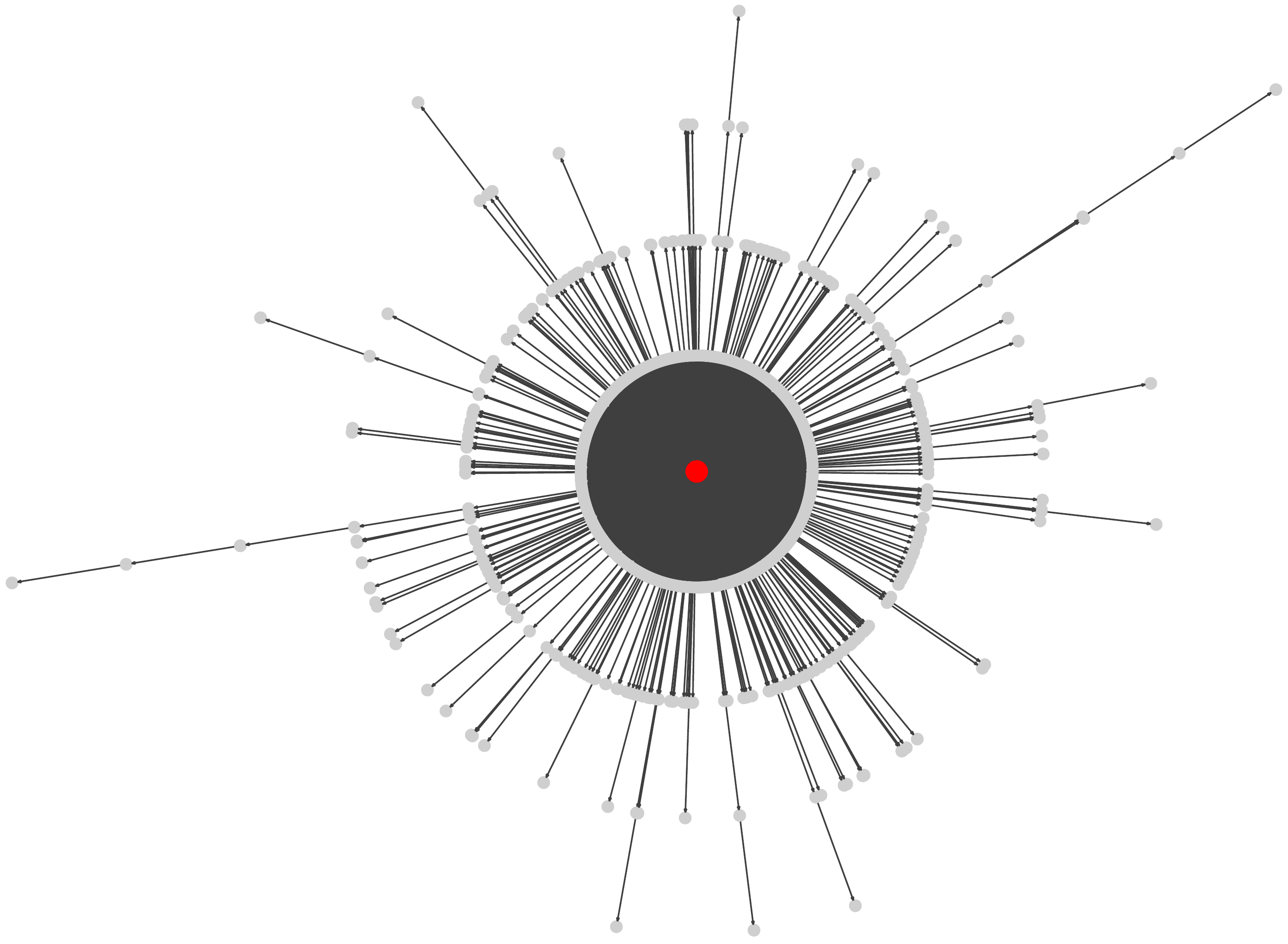}
	\caption{$d=2.47$}
	\end{subfigure}
	~ 
	\begin{subfigure}[b]{0.31\linewidth}
	\centering
	\includegraphics[width=0.6\linewidth]{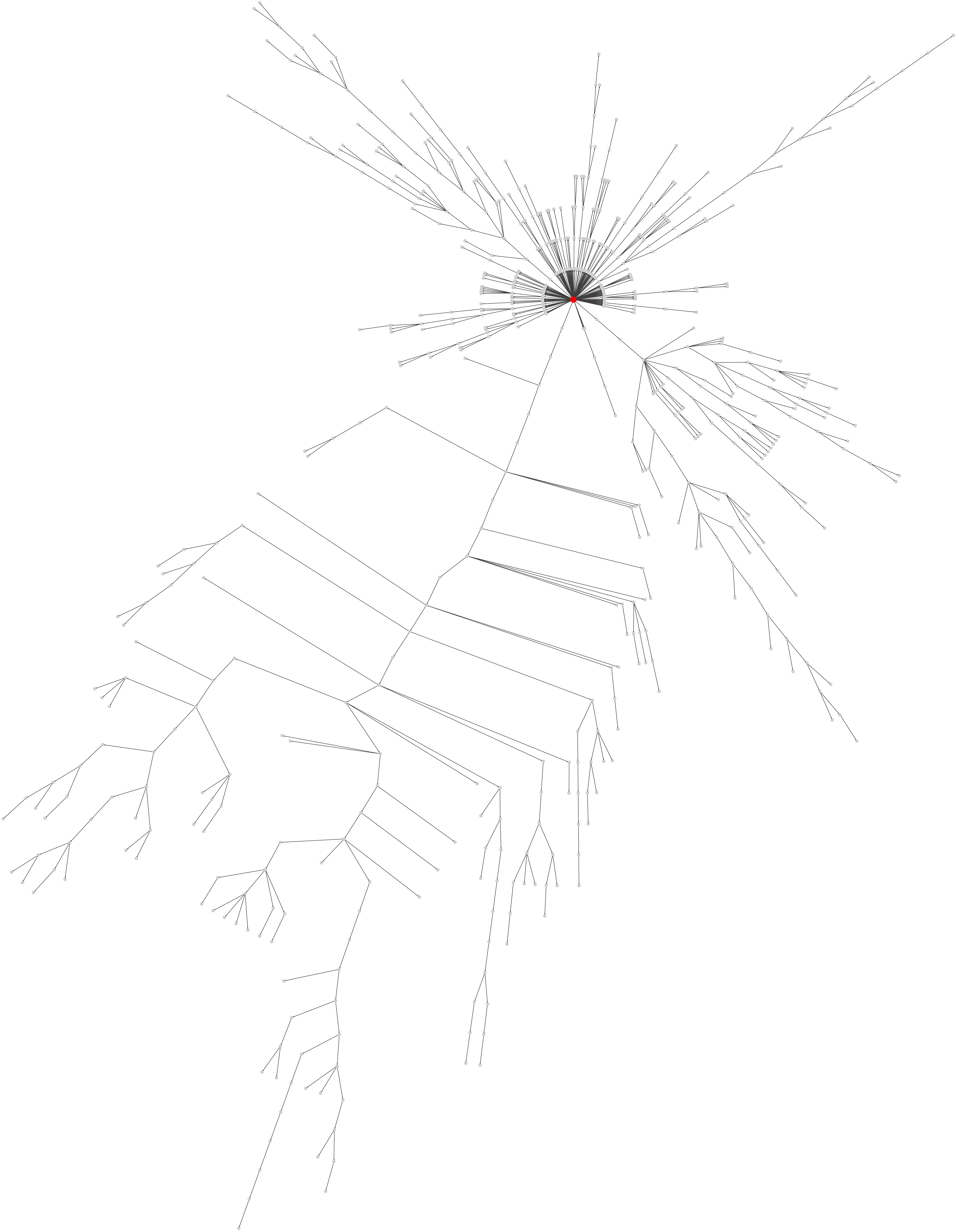}
	\caption{$d=14.4$}
	\end{subfigure}
	\vspace{-1mm}
	\caption{Cascades with a low Wiener index $d$ resemble star graphs, while those with a high index appear more viral (the root is red).}
	\label{fig:cascade_distances2}
	\vspace{-2mm}
\end{figure}

In addition to cascade growth, we quantify the shape of a cascade using the Wiener index, defined as the average distance between all pairs of nodes in a cascade. Recent work has proposed the Wiener index as a measure of the structural virality of a cascade~\cite{goelstructural}. Figure~\ref{fig:cascade_distances2} shows examples of cascades with varying Wiener index values. Intuitively, a cascade with low structural virality has most of its distribution following from a small number of hub nodes, while a cascade with high virality will have many long paths. Figure~\ref{fig:data} shows the distribution of cascade virality (as measured by Wiener index) in our dataset, which, as we saw with cascade size, follows a heavy-tailed distribution. While user cascades are typically smaller than page cascades in our dataset, they tend to have greater structural virality,
supporting the intuition that the structure of user-initiated
cascades is richer and deeper than that of page-initiated cascades.

\xhdr{Defining the cascade growth prediction problem}
Our aim in this paper is to study how well cascades can be predicted. Moreover, we are interested in understanding how various aspects of the prediction task affect the predictive performance.

There are several formulations of the task. If we were to define the task as a regression problem, predictions may be skewed towards large cascades, as cascade size follows a heavy-tailed distribution (Figure~\ref{fig:data}(right)).
Similarly, if we define it as a classification problem of predicting whether a cascade reaches a specific size, we may end up with unbalanced classes, and an overrepresentation of large cascades.
Also, if we simply observed a small initial portion of a cascade, and predict its future size, the problem is pathological as almost all cascades are small.
And, if we only varied the initial period of observation, the task of predicting whether a cascade reaches a certain size gets easier as we observe more of it.


To remedy these issues, we define a classification task that does not suffer from these deficiencies. We consider a binary classification problem where we observe the first $k$ reshares of a cascade and predict whether the eventual size of a cascade reaches the median size of all the cascades with at least $k$ reshares, $f(k)$.
This allows us to study how cascade predictibility varies with $k$.
As exactly half the cascades reach a size greater than the median by definition, random guessing achieves accuracy of 50\%.

Interestingly, the question of whether the cascade will reach $f(k)$ is equivalent to that of whether a cascade will double in size. This follows directly from the fact that cascade size distribution follows a power-law with exponent $\alpha \approx 2$. Consider a power-law distribution on the interval $(x_{min}, \infty)$ with a power-law exponent $\alpha \approx 2$. Then the median $f(x)$ of this distribution is $2 \cdot x_{min}$, as demonstrated by the following calculation:
\begin{displaymath}
\int_{x_{\min}}^{f(x)} \frac{\alpha-1}{x_{\min}}\left(\frac{x}{x_{\min}}\right)^{-\alpha} dx = \frac{1}{2} \Rightarrow f(x) = 2^{\frac{1}{\alpha-1}}x_{\min} = 2x_{\min}
\end{displaymath}

As we examine cascades of size greater than $k=x_{min}$, the median size of these cascades is thus $2\cdot k$ from this derivation. In each of our prediction tasks, we observe that this is indeed true.



\xhdr{Methods used for learning}
Our general methodology for the cascade prediction problem will be to represent a cascade by a set of features and then use machine learning classifiers to predict its future size.
We used a variety of learning methods, including linear regression, naive Bayes, SVM, decision trees and random forests. However, we primarily report performance of the logistic regression classifier for ease of comparison. In many cases, the performance of most classifiers was similar, although non-linear classifiers such as random forests usually performed slightly better than linear classifiers such as logistic regression. In all cases, we performed 10-fold cross validation and report the classification accuracy, F1 score, and area under the ROC curve (AUC).

\subsection{Factors driving cascade growth}
\label{sec:features}

We proceed by describing factors that contribute to the growth and spreading of cascades. We group these factors into five classes: properties of the content that is spreading, features of the original poster, features of the resharer, structural features of the cascade, and temporal characteristics of the cascade. Table \ref{tab:featurelist} contains a detailed list of features.

\xhdr{Content features}
%
The first natural factor contributing to the ability of the cascade to spread is the content itself~\cite{berger12viral}. 
On Twitter, tweet content and in particular, hashtags, are used to generate content features \cite{ma2013predicting,tsur2012s}, and identify topics affecting retweet likelihood \cite{petrovic2011rt}.
LDA topic models have also been incorporated into these prediction tasks \cite{Hong2011www}, and human raters employed to infer the interestingness of content \cite{bakshy2011everyone,petrovic2011rt}.
In our work, we relied on a linear SVM model, trained using image GIST descriptors and color histogram features, to assign likelihood scores of a photo being a closeup shot, taken indoors or outdoors, synthetically generated (e.g., screenshots or pure text vs.\ photographs), or contained food, a landmark, person, nature, water, or overlaid text (e.g., a meme).
We also analyzed words in the caption accompanying an image for positive sentiment, negative sentiment, and sociality~\cite{jenders13analyzing,pennebaker2001linguistic}.

Nevertheless, while content features affected the performance of structural and temporal features, we find that they are weak predictors of how widely disseminated a piece of content would become.


\xhdr{Original poster/resharer features} 
Some prior work focused on features of the root note in a cascade to predicting the cascade's evolution, finding that content from highly-connected individuals reaches larger audiences, and thus spreads further.
Users with large follower counts on Twitter generated the largest retweet cascades~\cite{bakshy2011everyone}.
Separately, features of an author of a tweet were shown to be more important than features of the tweet itself~\cite{petrovic2011rt}.
In many Twitter studies predicting cascade size or popularity, a user's number of followers ranks among the top, if not the most, important predictor of popularity~\cite{bakshy2011everyone, ma2013predicting}.

Other features of the root node have also been studied, such as the number of prior retweets of a user's posts~\cite{bakshy2011everyone,Hong2011www}, and how many Twitter lists a user was included in~\cite{petrovic2011rt}.
The number of @-mentions of a Twitter user was used to predict whether, and how soon a tweet would be retweeted, how many users would directly retweet, and the depth a cascade would reach~\cite{yang2010predicting}.
Still, \cite{cha2010measuring} found that various measures of a user's popularity are not very correlated with his or her influence.

We capture the intuition behind these factors by defining demographic as well as network features of the original poster as well as the features of the users who reshared the content so far.
We use Facebook's distinction of users (individuals) and pages (entities representing an interest) to further distinguish different origin types, in addition to the influence features mentioned above.


\xhdr{Structural features of the cascade}
Networks provide the substrate through which information spreads, and thus their structure influences the path and reach of the cascade.
As illustrated in Figure \ref{fig:cascade}, we generate features from both the graph of the first $k$ reshares ($\hat G$), as well as the induced friend subgraph of the first $k$ resharers ($G'$).
Whereas the reshare graph $\hat G$ describes the actual spread of a cascade, the friend subgraph $G'$ provides information about the social ties between these initial resharers. The social graph $G$ allows us to compute the potential reach of these reshares.

Previous work considered the network structure of the underlying graph in inferring the virality of content~\cite{weng2013virality}, with highly viral items spreading across communities.
We use the density of the initial reshare cascade ($\mathit{subgraph}'_k$) and the proximity to the root node ($\mathit{orig\_connections}_k$, $\mathit{did\_leave}$) as proxies for whether an item is spreading primarily within a community or across many.
One can also look outside the network between resharers, and count the number of users reachable via all friendship and follow edges of the first $k$ users ($\mathit{border\_nodes}_k$). This relates to total number of exposed users, and has been demonstrated to be an important feature in predicting Twitter hashtag popularity \cite{ma2013predicting}.

As we can trace information flow on Facebook exactly, we need not worry about independent entry points influencing a cascade \cite{bakshy2009social,myers2010external}; external influence instead allows us to investigate multiple independent cascades arising from the same content (see Section~\ref{sec:content_control}).


\xhdr{Temporal features}
Properties related to the ``speed'' of the cascade (e.g., $\mathit{time}_k$) were shown to be the most important features in predicting thread length on Facebook \cite{backstrom2013characterizing}, and are a primary mechanism in predicting online content popularity \cite{szabo2010predicting}.
Moreover, as the speed of diffusion changes over time, this may have a strong effect on the ability of the cascade to continue spreading through the network~\cite{yang2010predicting}.

We characterize a number of temporal properties of cascade diffusion (see Table \ref{tab:featurelist}). In particular, we measure the change in the speed of reshares ($\mathit{time}''_{1..k}$), compare the differences between the speed in the first and second half of the measurement period ($\mathit{time}'_{1..k/2}$, $\mathit{time}'_{k/2..k}$), and qunatify the number of users who were exposed to the cascade per time unit ($\mathit{views}'_{1..k-1,~k}$).

\begin{table*}[!ht]
\small
\centering
\ra{1.3}
\begin{tabular}{@{}ll@{}}\toprule
\multicolumn{2}{c}{\textbf{Content Features}} \\
$\mathit{score}_{\mathit{food}/\mathit{nature}/\mathit{\ldots}}$ & The probability of the photo having a specific feature (food, overlaid text, landmark, nature, etc.) \\ 
$\mathit{is\_en}$ & Whether the photo was posted by an English-speaking user or page \\ 
$\mathit{has\_caption}$ & Whether the photo was posted with a caption \\ 
$\mathit{liwc}_{\mathit{pos}/\mathit{neg}/\mathit{soc}}$ & Proportion of words in the caption that expressed positive or negative emotion, or sociality, if English \\ 
\midrule
\multicolumn{2}{c}{\textbf{Root (Original Poster) Features}} \\
$\mathit{views}_{0,~k}$ & Number of users who saw the original photo until the $k$th reshare was posted\\ 
$\mathit{orig\_is\_page}$ & Whether the original poster is a page \\ 
$\mathit{outdeg}(v_0)$ & Friend, subscriber or fan count of the original poster\\ 
$\mathit{age}_0$ & Age of the original poster, if a user \\ 
$\mathit{gender}_0$ & Gender of the original poster, if a user \\ 
$\mathit{fb\_age}_0$ & Time since the original poster registered on Facebook, if a user \\ 
$\mathit{activity}_0$ & Average number of days the original poster was active in the past month, if a user \\ 
\midrule
\multicolumn{2}{c}{\textbf{Resharer Features}} \\
$\mathit{views}_{1..k-1,~k}$ & Number of users who saw the first $k-1$ reshares until the $k$th reshare was posted\\ 
$\mathit{pages}_k$ & Number of pages responsible for the first $k$ reshares, including the root, or $\sum_{i=0}^k \mathds{1}\{\text{$v_i$ is a page}\}$\\ 
$\mathit{friends}_k^{\mathit{avg}/\mathit{90p}}$ & Average or 90th percentile friend count of the first $k$ resharers, or $\frac{1}{k}\sum_{i=1}^k \mathit{outdeg}_{\mathit{friends}}(v_i) \mathds{1}\{\text{$v_i$ is a user}\}$\\ 
$\mathit{fans}_k^{\mathit{avg}/\mathit{90p}}$ & Average or 90th percentile fan count of the first $k$ resharers, or $\frac{1}{k}\sum_{i=1}^k \mathit{outdeg}(v_i) \mathds{1}\{\text{$v_i$ is a page}\}$\\ 
$\mathit{subscribers}_k^{\mathit{avg}/\mathit{90p}}$ & Average or 90th percentile subscriber count of the first $k$ resharers, or $\frac{1}{k}\sum_{i=1}^k \mathit{outdeg}_{\mathit{subscriber}}(v_i) \mathds{1}\{\text{$v_i$ is a user}\}$\\ 
$\mathit{fb\_ages}_k^{\mathit{avg}/\mathit{90p}}$ & Average or 90th percentile time since the first $k$ resharers registered on Facebook, or $\frac{1}{k}\sum_{i=1}^k \mathit{fb\_age}_i$\\ 
$\mathit{activities}_k^{\mathit{avg}/\mathit{90p}}$ & Average number of days the first $k$ resharers were active in July, or $\frac{1}{k}\sum_{i=1}^k \mathit{activity}_i$\\ 
$\mathit{ages}_k^{\mathit{avg}/\mathit{90p}}$ & Average age of the first $k$ resharers, or $\frac{1}{k}\sum_{i=1}^k \mathit{age}_i$\\ 
$\mathit{female}_k$ & Number of female users among the first $k$ resharers, or $\sum_{i=1}^k \mathds{1}\{\mathit{gender}_i\text{ is female}\}$\\ 
\midrule
\multicolumn{2}{c}{\textbf{Structural Features}} \\ 
$\mathit{outdeg}(v_i)$ & Connection count (sum of friend, subscriber and fan counts) of the $i$th resharer (or out-degree of $v_i$ on $G=(V,E)$) \\ 
$\mathit{outdeg}(v'_i)$ & Out-degree of the $i$th reshare on the induced subgraph $G'=(V',E')$ of the first $k$ resharers and the root \\ 
$\mathit{outdeg}(\hat{v}_i)$ & Out-degree of the $i$th reshare on the reshare graph $\hat G = (\hat V, \hat E)$ of the first $k$ reshares \\ 
$\mathit{orig\_connections}_k$ & Number of first $k$ resharers who are friends with, or fans of the root, or $|\{v_i~|~(v_0, v_i) \in E,~1 \le i \le k\}|$\\ 
$\mathit{border\_nodes}_k$ & Total number of users or pages reachable from the first $k$ resharers and the root, or $|\{v_i~|~(v_i, v_j) \in E,~0 \le i,j \le k\}|$\\ 
$\mathit{border\_edges}_k$ & Total number of first-degree connections of the first $k$ resharers and the root, or $|\{(v_i, v_j)~|~(v_i, v_j) \in E,~0 \le i,j \le k\}|$\\ 
$\mathit{subgraph}'_k$ & Number of edges on the induced subgraph of the first $k$ resharers and the root, or $|\{(v_i, v_j)~|~(v_i, v_j) \in E',~0 \le i,j \le k\}|$\\ 
$\mathit{depth}'_k$ & Change in tree depth of the first $k$ reshares, or $\min_{\beta} \sum_{i=1}^k (\mathit{depth}_i - \beta i)^2$\\ 
$\mathit{depths}_k^{\mathit{avg}/\mathit{90p}}$ & Average or 90th percentile tree depth of the first $k$ reshares, or $\frac{1}{k} \sum_{i=1}^k \mathit{depth}_i$ \\ 
$\mathit{did\_leave}$& Whether any of the first $k$ reshares are not first-degree connections of the root \\ 
\midrule
\multicolumn{2}{c}{\textbf{Temporal Features}} \\
$\mathit{time}_i$ & Time elapsed between the original post and the $i$th reshare \\ 
$\mathit{time}'_{1..k/2}$ & Average time between reshares, for the first $k/2$ reshares, or $\frac{1}{k/2-1}\sum_{i=1}^{k/2-1} (\mathit{time}_{i+1} - \mathit{time}_i)$ \\ 
$\mathit{time}'_{k/2..k}$ & Average time between reshares, for the last $k/2$ reshares, or $\frac{1}{k/2-1}\sum_{i=k/2}^{k-1} (\mathit{time}_{i+1} - \mathit{time}_i)$ \\ 
$\mathit{time}''_{1..k}$ & Change in the time between reshares of the first $k$ reshares, or $\min_{\beta} \sum_{i=1}^{k-1} (\mathit{time}_{i+1} - \mathit{time}_i) - \beta i)^2$\\ 
$\mathit{views}'_{0,k}$ & Number of users who saw the original photo, until the $k$th reshare was posted, per unit time, or $\frac{\mathit{views}_{0,~k}}{\mathit{time}_k}$\\ 
$\mathit{views}'_{1..k-1,~k}$ & Number of users who saw the first $k-1$ reshares, until the $k$th reshare was posted, per unit time, or $\frac{\mathit{views}_{1..k-1,k}}{\mathit{time}_k}$\\ 
\bottomrule
\end{tabular}
\caption{List of features used for learning. We compute these features given the cascade until the $k$th reshare.}
\vspace{-3mm}
\label{tab:featurelist}
\end{table*}

\subsection{Predicting cascade growth}

\begin{figure}[t]
\centering
\includegraphics[width=\linewidth]{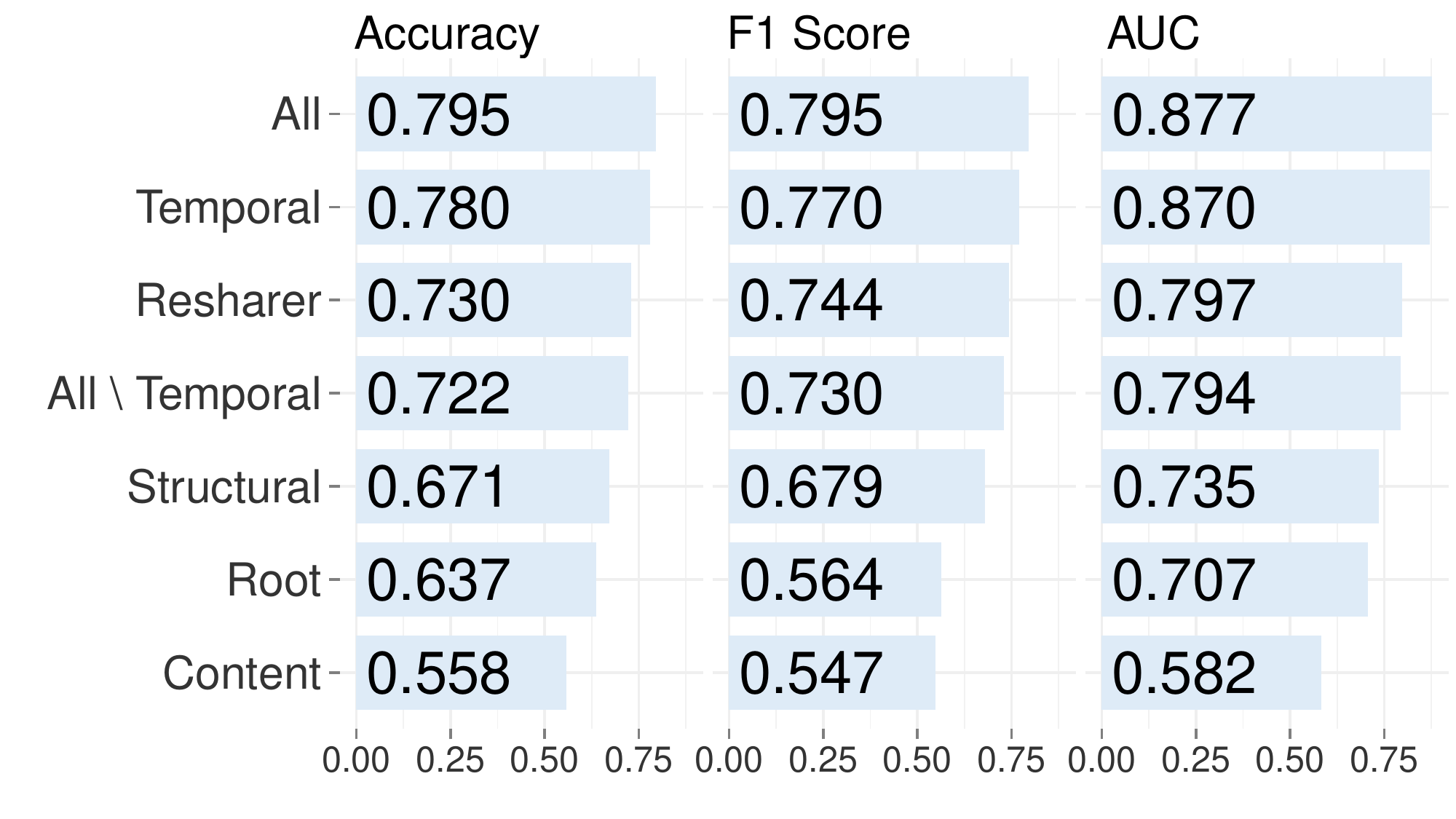}
\caption{Using logistic regression, we are able to predict with near 80\% accuracy whether the size of a cascade will reach the median (10) after observing the first $k=$5 reshares.}
\label{fig:general_prediction}
\vspace{-0.5em}
\end{figure}

To illustrate the general performance of the features described in the previous section we consider a simple prediction task, where we observe the first 5 reshares of the cascade and want to predict whether it will reach the median cascade size (or equivalently, whether it will double and be reshared at least 10 times). For the experiment we use a set of $N_c =$ 150,572 photos, where each photo was shared at least 5 times. The total number of reshares of these photos was $N_r =$ 9,233,300.

Figure \ref{fig:general_prediction} shows logistic regression performance using all features from Table~\ref{tab:featurelist}. For this task, random guessing would obtain a performance of 0.5, while our method achieves surprisingly strong performance: classification accuracy of 0.795 and AUC of 0.877.
If we relax the task and instead of predicting above vs.\ below median size, we predict top vs.\ bottom quartile (top 25\% vs.\ bottom 25\%) the accuracy rises even further to 0.926, and the AUC to 0.976.

Overall, while each feature set is individually significantly better than predicting at random, it is the set of temporal features that outperforms all other individual feature sets, obtaining performance scores within 0.025 of those obtained when using all features. To understand if we could do well without temporal features, we trained a classifier which excluded them and were still able to obtain reasonable performance even without these features. This is especially useful when one knows through \emph{whom} information was passed, but not \emph{when} it was passed. The lack of reliance on any individual set of features demonstrates that the predictions are robust.

Studied individually, we also find that temporal features generally performed best, followed by structural features.
The reshare rate in the second half ($\mathit{time}'_{k/2..k}$) was most predictive, attaining accuracy of 0.73.
This was followed by the rate of user views of the original photo, $\mathit{views}'_{0,k}$, and the time elapsed between the original post and fifth reshare, $\mathit{time}_5$ (both 0.72). In fact, $\mathit{time}_{k+1}$ is always more accurate than $\mathit{time}_{k}$.
The most accurate structural features were $\mathit{did\_leave}$ and $\mathit{outdeg}(v_0)$ (both 0.65).
We examine individual feature importance in more detail later. 




\subsection{Predictability and the observation window of size $k$}

\begin{figure}[t]
\centering
\includegraphics[width=\linewidth]{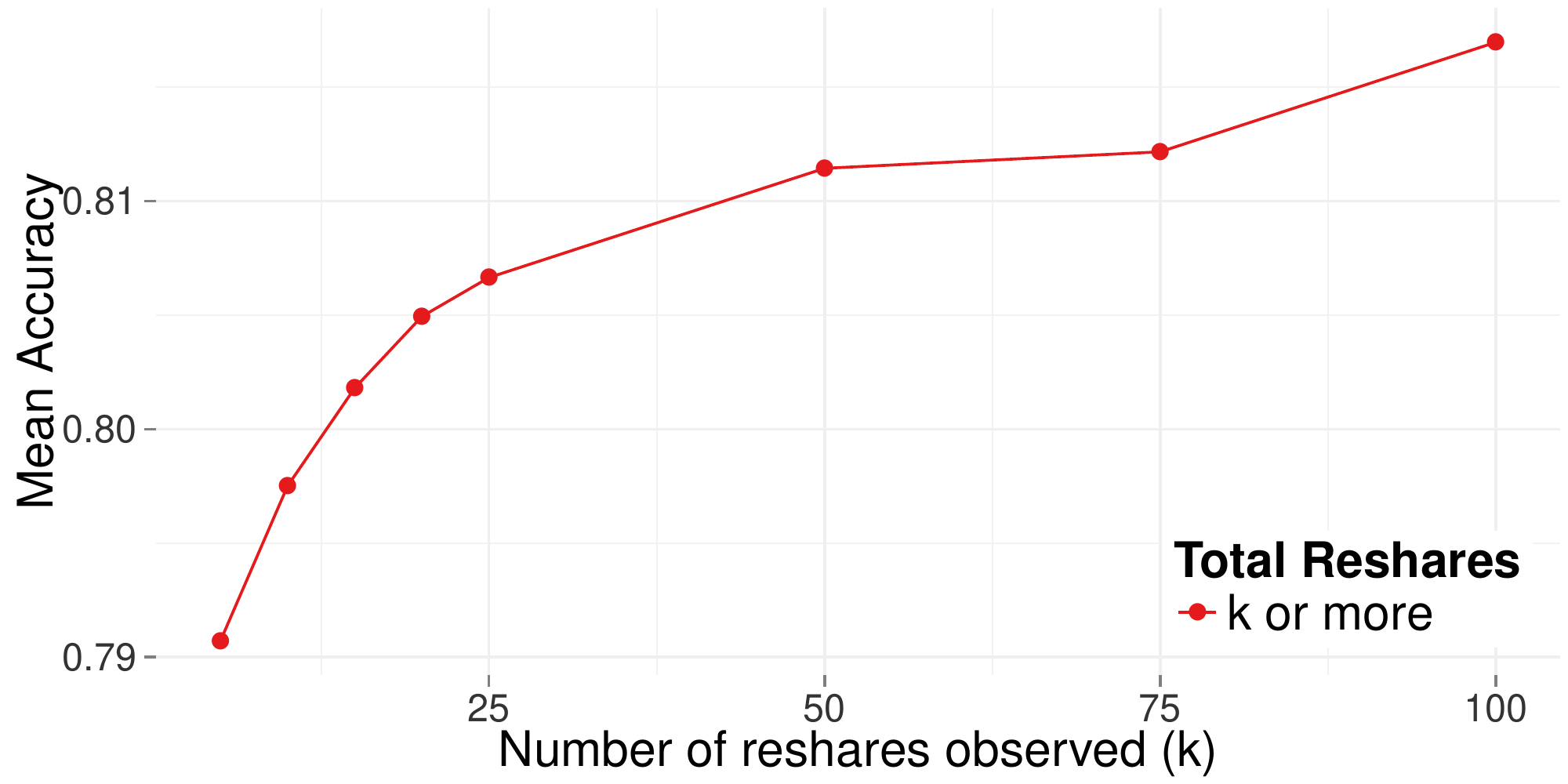}
\caption{If we observe the first $k$ reshares of a cascade, and want to predict whether the cascade will double in size, our prediction improves as we observe more of it.}
\label{fig:varyk_prediction_base}
\vspace{-0.5em}
\end{figure}

It is also natural to ask whether cascades get more or less predictable as we observe more of the initial growth of a cascade. One may think that observing more of the cascade may allow us to extrapolate its future growth better; on the other hand, additional observed reshares may also introduce noise and uncertainty in the future growth of the cascade. Note that the task does not get easier as we observe more of the cascade, as we are predicting whether the cascade will reach size $2k$ (or equivalently, the median) given that we have seen $k$ reshares so far.

Figure \ref{fig:varyk_prediction_base} shows that the predictive performance of whether a cascade doubles in size increases as a function of the number of observed reshares $k$.
In other words, it is easier to predict whether a cascade that has reached 25 reshares will get another 25, than to predict whether a cascade that has reached 5 reshares will obtain an additional 5. Thus, the prediction accuracy for larger cascades is above the already high accuracy for smaller values of $k$. The change in the F1 score and AUC also follow a very similar trend.

Overall, these results demonstrate that observing more of the cascade, while also predicting ``farther'' into the future, is easier than observing a cascade early in its life and predicting what it will do next (i.e., $k=5$ vs. $k=25$).


\begin{figure}[t]
\centering
\includegraphics[width=\linewidth]{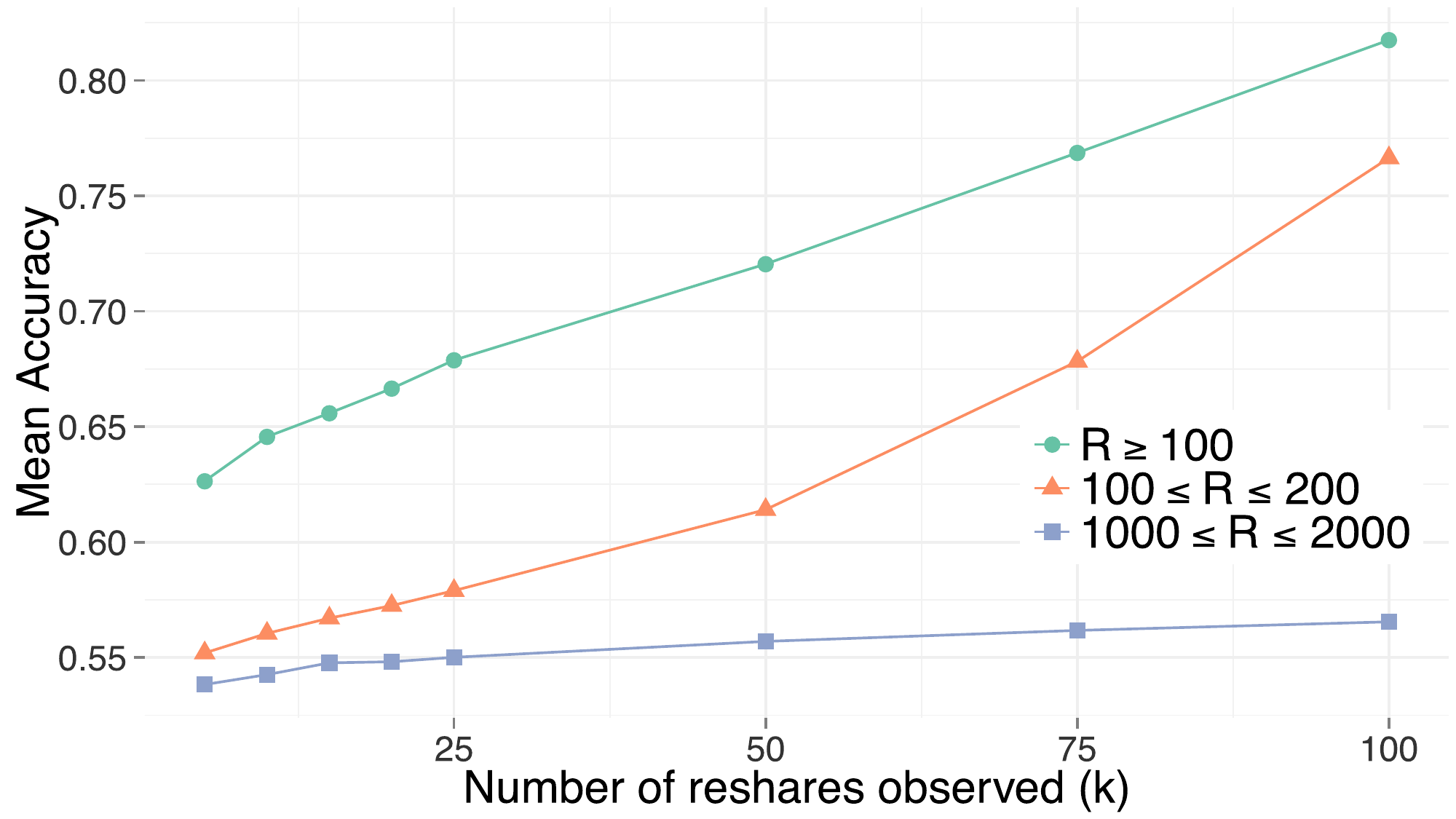}
\caption{Knowing that a cascade obtains at least $R$ reshares, prediction performance increases linearly with $k$, $k \le R$. However, differentiating among cascades with large $R$ also becomes more difficult.}
\label{fig:varyk_prediction}
\vspace{-3mm}
\end{figure}

\xhdr{Fixing the minimum cascade size $R$}
In the previous version of the task, cascades are required only to have at least $k$ reshares.
Thus, the set of cascades changes with $k$.
Here, we examine a variation of this task, where we compose a dataset of cascades that have at least $R$ reshares. 
We observe the first $k$ ($k \le R$) reshares of the cascade and aim to predict whether the cascade will grow over the median size (over all cascades of size $\ge R$).
As we increase $k$, the task gets easier as we observe more of the cascade and the predicted quantity does not change.

With the task, we find that performance increases linearly with $k$ up to $R$, or that there is no ``sweet spot'' or region of diminishing returns ($p < $ 0.05 using a Harvey-Collier test).
For example, the top-most line in Figure \ref{fig:varyk_prediction} shows that when each observed cascade has obtained 100 or more reshares, performance increases linearly as more of the cascade is observed.
This demonstrates that more information is always better: the greater the number of observed reshares, the better the prediction.

However, Figure \ref{fig:varyk_prediction} also shows that larger cascades are less predictable than smaller cascades.
For example, predicting whether cascades with 1,000 to 2,000 reshares grow large is significantly more difficult than predicting cascades of 100 to 200 reshares.
This shows that once one knows that a cascade will grow to be large, knowing the characteristics of the very beginning of its spread is less useful for prediction.



\begin{figure}[ht!]
  \centering
  \begin{subfigure}[b]{1\linewidth}
  \centering
  \caption{Content}
  \includegraphics[width=1\linewidth]{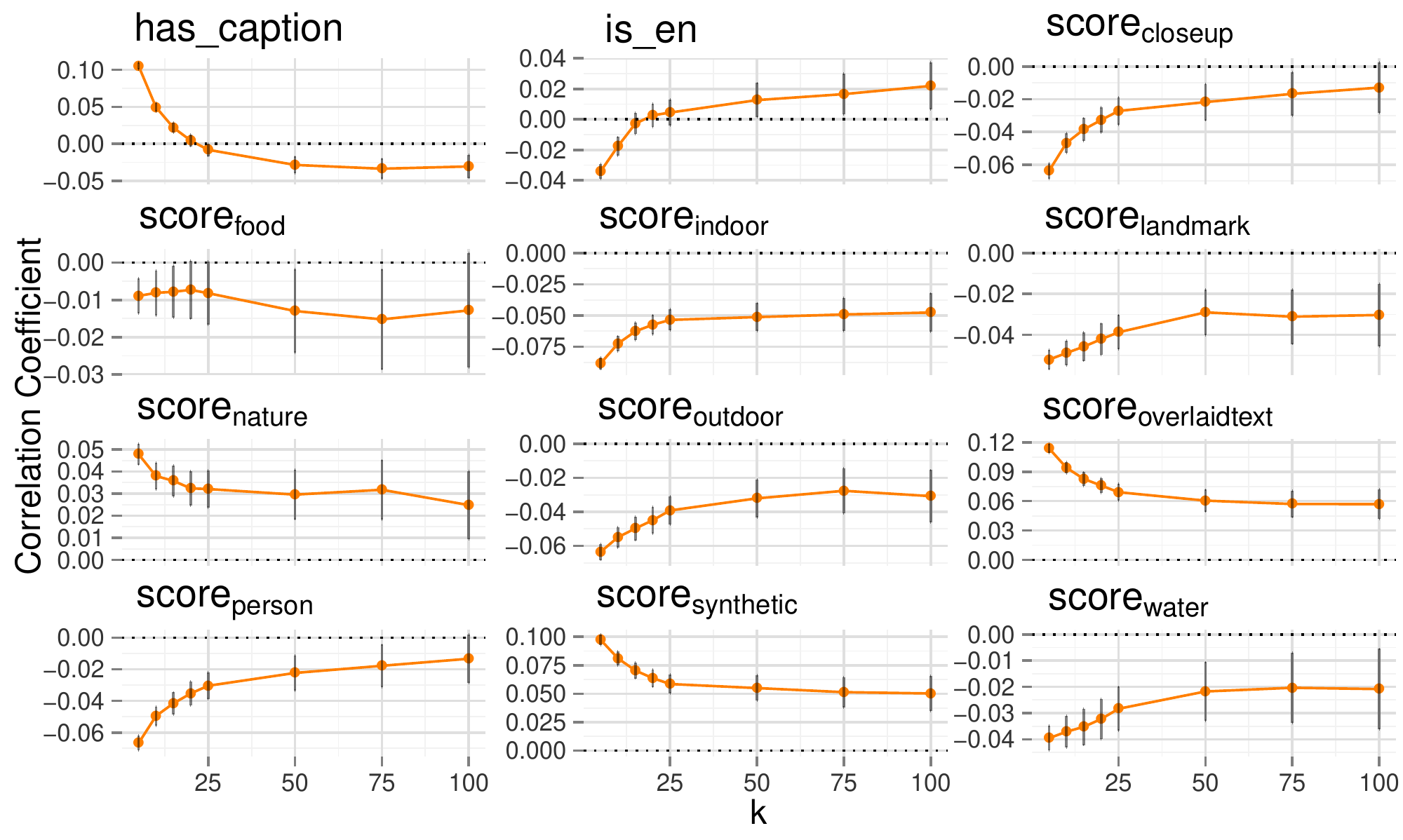}
  \label{fig:varyk_content}
  \end{subfigure}

  \vspace{-1em}

  \begin{subfigure}[b]{1\linewidth}
  \centering
  \caption{Root}
  \includegraphics[width=1\linewidth]{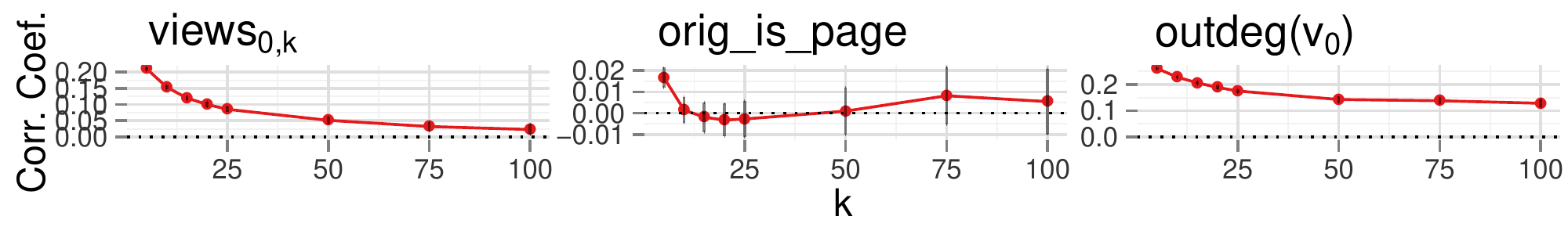}
  \label{fig:varyk_root}
  \end{subfigure}

  \vspace{-1em}

  \begin{subfigure}[b]{1\linewidth}
  \centering
  \caption{Resharer}
  \includegraphics[width=1\linewidth]{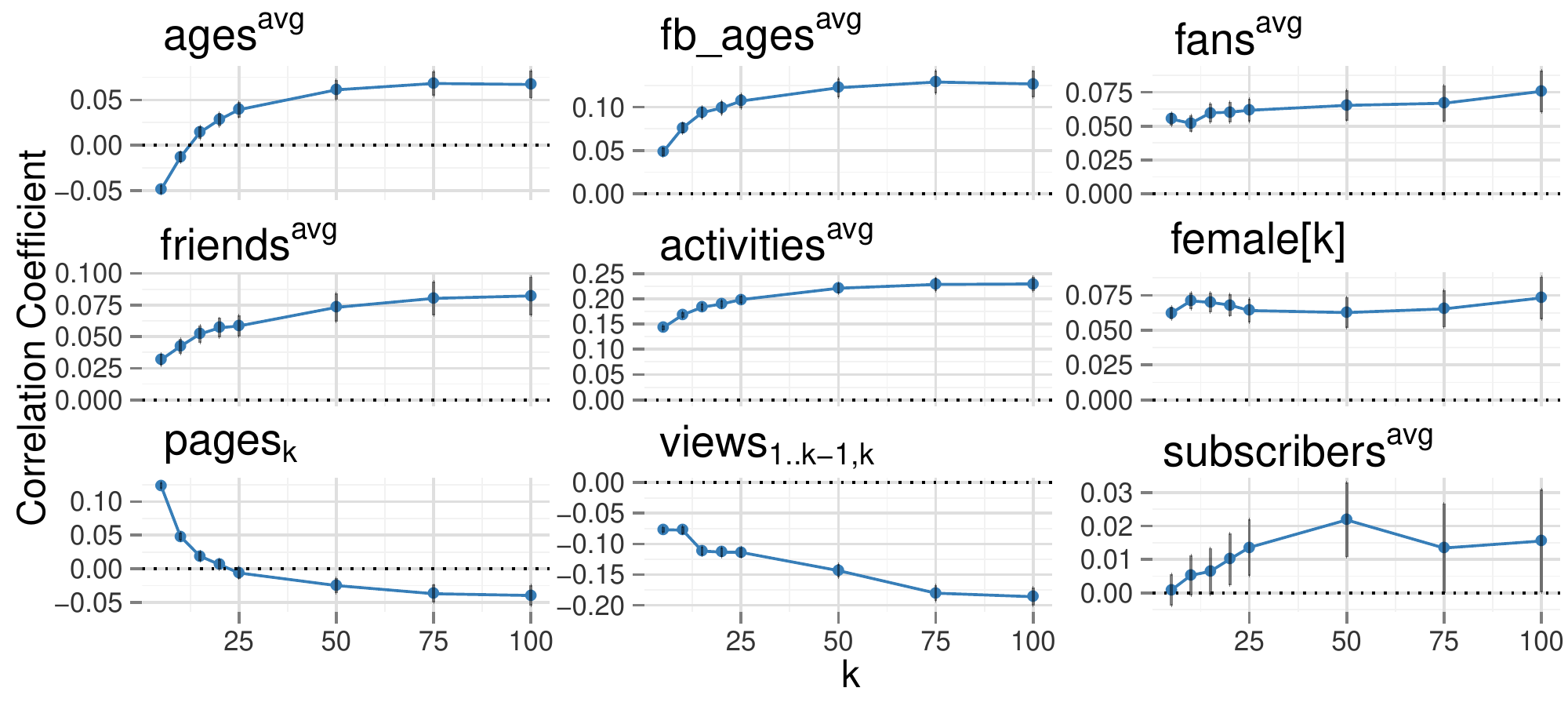}
  \label{fig:varyk_resharer}
  \end{subfigure}

  \vspace{-1em}

  \begin{subfigure}[b]{1\linewidth}
  \centering
  \caption{Structural}
  \includegraphics[width=1\linewidth]{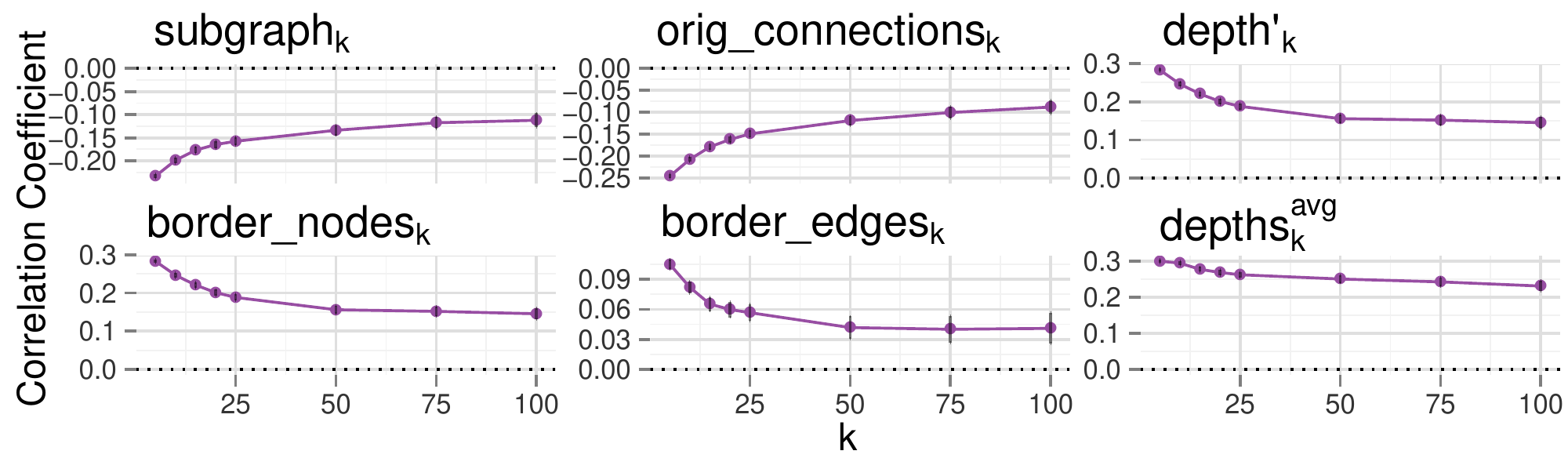}
  \label{fig:varyk_structural}
  \end{subfigure}

  \vspace{-1em}

  \begin{subfigure}[b]{1\linewidth}
  \centering
  \caption{Temporal}
  \includegraphics[width=1\linewidth]{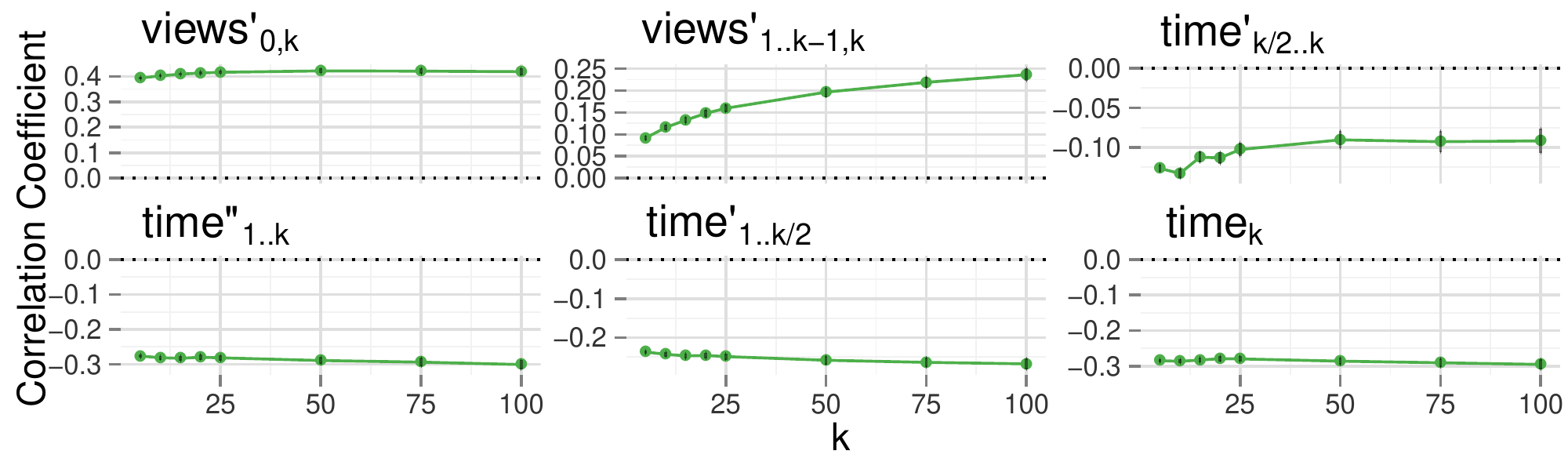}
  \label{fig:varyk_temporal}
  \end{subfigure}
  \vspace{-3.0em}
  \caption{The importance of each feature varies as we observe more of a cascade, as shown by the change in correlation coefficients.}
  \label{fig:varyk}
  \vspace{-1em}
\end{figure}

\subsection{Changes in feature importance}

We now examine how feature importance changes as more and more of the cascade is observed. In this experiment, we compute the value of the feature after observing first $k$ reshares and measure the correlation coefficient of the feature value with the log-transformed number of reshares (or cascade size).


Figure~\ref{fig:varyk} shows the results for the five feature types. We summarize the results by the following observations:
\begin{itemize}
	\item \emph{Correlations of averages increase with the number of observations}. As we obtain more examples, naturally averages get less noisy, and more predictive (e.g., $\mathit{ages}^\mathit{avg}$ and $\mathit{friends}^\mathit{avg}$). \item \emph{The original post gets less important with increasing $k$}. After observing 100 reshares, it becomes less important that the original post was made by a page ($\mathit{orig\_is\_page}$), or that the original poster had many connections to other users ($\mathit{outdeg}(v_0)$).
	\item \emph{Similarly, the actual content being reshared gets less important with increasing $k$}. Almost all content features tend to zero as $k$ increases, except for $\mathit{has\_caption}$ and $\mathit{is\_en}$.
	This can be explained by the fact that cascades of photos with captions have a unimodal distribution, and cascades started by English speakers have a bimodal distribution. Thus, these features become correlated in opposite directions.
	\item \emph{Successful cascades get many views in a short amount of time, and achieve high conversion rates}. The number of users who have viewed reshares of a cascade is more negatively correlated with increasing $k$ ($\mathit{views}_{1..k-1,k}$), suggesting that requiring ``fewer tries'' to achieve a given number of reshares is a positive indicator of its future success. On the other hand, while requiring fewer views is good, rapid exposure, or reaching many users within a short amount of time is also a positive predictor ($\mathit{views}'_{1..k-1,k}$).
	\item \emph{Structural connectedness is important, but gets less important over time}. Nevertheless, reshare depth remains highly correlated: the deeper a cascade goes, the more likely it is to be long-lasting, as even users ``far away'' from the original poster still find the content interesting.
	\item \emph{The importance of timing features remains relatively stable}. While highly correlated, timing features remain remarkably stable in importance as $k$ increases.
\end{itemize}

We note individual features' logistic regression coefficients empirically follow similar shapes, but have the downside of having interactions with one another. Using either the slope of the best-fit line of the cascade size against the normalized feature value, or individual feature performance also reveals similar trends. Further LIWC text content features (positive, negative, and social categories) consistently performed poorly, attaining performance no better than chance, with accuracy between 0.49 and 0.52.

\section{Predicting Cascade Structure}
\label{sec:structure}

Similar to predicting cascade size we can also attempt to predict the {\em structure} of the cascade.
We now turn to examining how structural features of the cascade determine its evolution and spread.

\subsection{User-started and page-started cascades}

\begin{figure}[t]
  \centering
  \includegraphics[width=\linewidth]{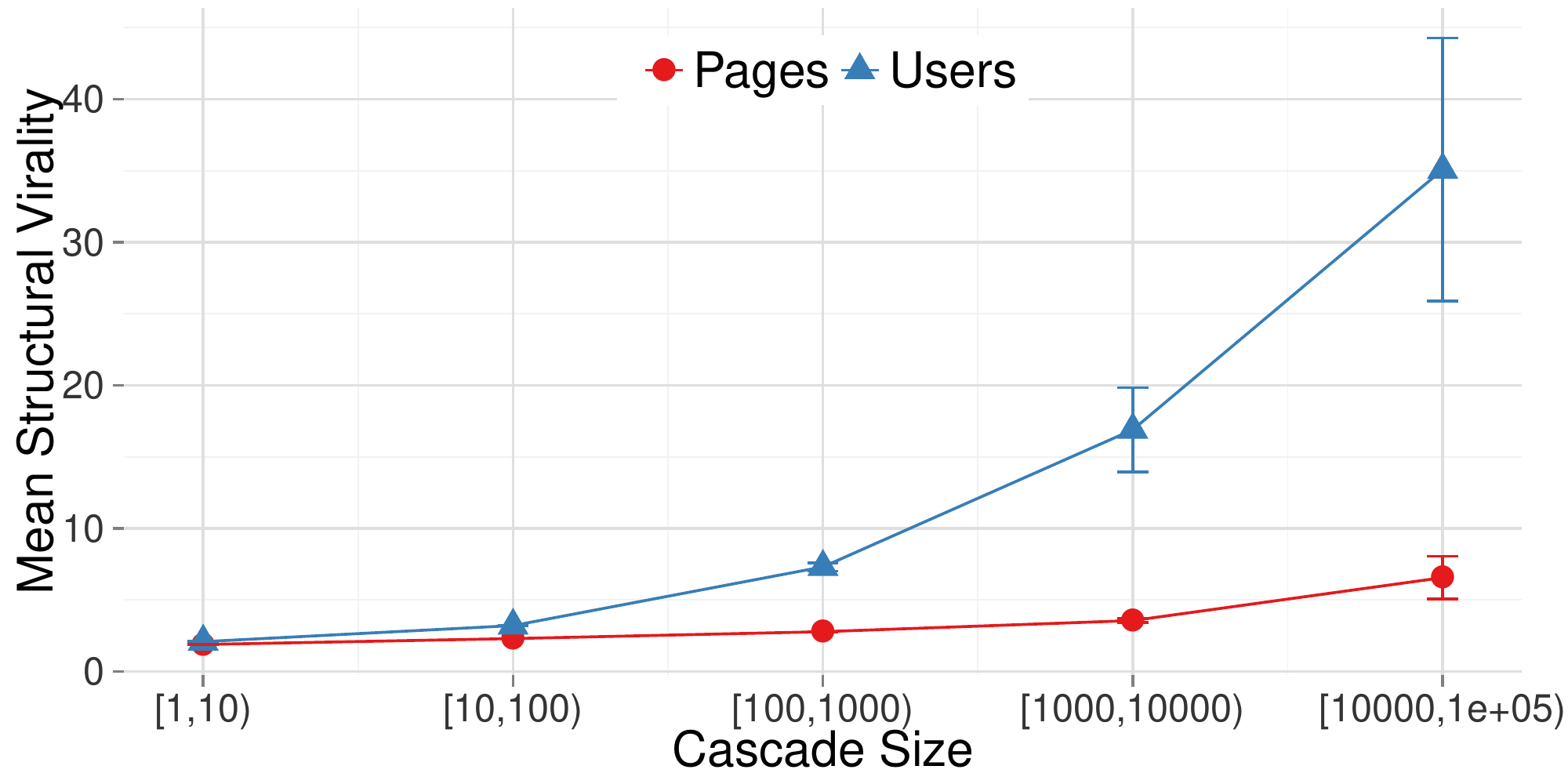}
  \caption{The mean structural virality (Wiener index) increases with cascade size, but is significantly higher for user cascades.}
\label{fig:viralitybysize}
\vspace{-0.5em}
\end{figure}
Earlier we discussed the notion of {\em structural virality} as
a measure of how much the structure of a cascade is dominated by
a few hub nodes, and we saw that user-initiated cascades have
significantly higher structural virality than page-initiated cascades,
reflecting their richer graph structure. 
It is natural to ask how these distinctions vary with the size of the cascade ---
are large user-initiated cascades more similar to page-initiated ones, e.g. are
they driven by popular hub nodes?

Figure~\ref{fig:viralitybysize} shows that the opposite is the 
case --- user and page-initiated cascades remain structurally distinct, with this distinction even increasing with cascade size.
Moreover, this difference continues to hold even when controlling for the number of first-degree reshares (directly from the root),
suggesting a certain robustness to their richer structure.
Because of these structural differences, we handle user and page cascades separately in the analyses that follow.


These distinctions may also help explain a large 
difference in the predictability of
user-initiated vs.\ page-initiated cascades. 
We observe that
for page cascades accuracy exceeds 80\%, while that for user cascades
is slightly under 70\%. (These results also hold for the F1 score and
AUC, with a difference of about 0.1.)
The fact that much more of the structure of a page-initiated cascade
is typically carried by a small number of hub nodes may suggest
why the prediction task is more tractable in this case.

\subsection{The initial structure of a cascade influences its eventual size}

\begin{figure}[t]
	\centering
	\includegraphics[width=\linewidth]{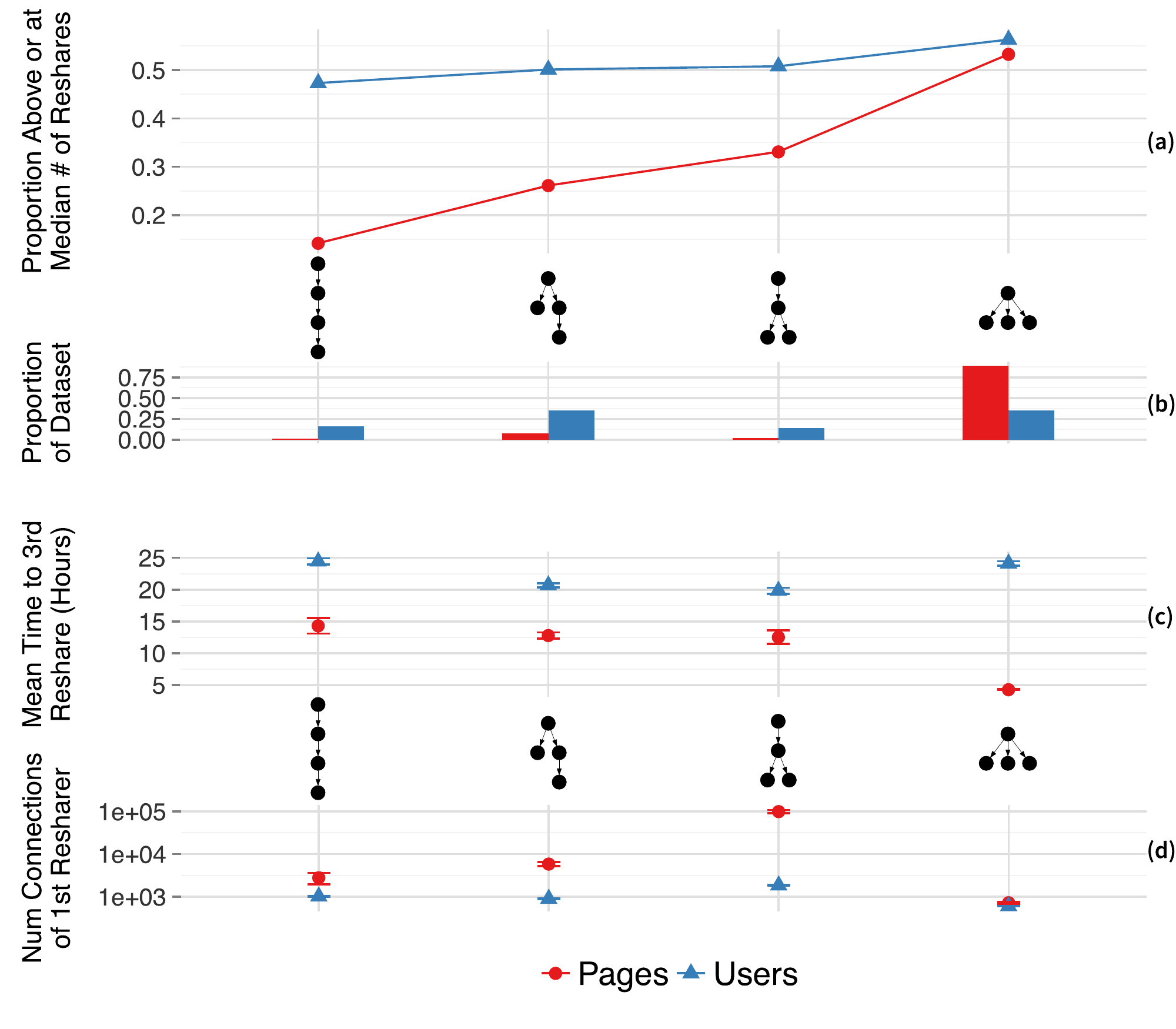}
	\caption{Shallow initial cascade structures are indicative of larger cascades. In contrast to page-started cascades, where the mean time to the 3rd reshare decreases with decreasing depth of the initial cascade, shallow cascades take a much longer time to form for user-started cascades. For these, the connections of the 1st resharer also significantly impacts the time to the 3rd resharer, especially when it receives two reshares before the original receives a second.}
\label{fig:paper_shapes3}
\vspace{-0.5em}
\end{figure}

To understand how structure bears on the future growth of the cascade, we examine how the configuration of the first three reshares (and the root) correlates with the cascade size.
In particular, we measure the proportion of cascades starting from each configuration that reach the median size. We do this separately for two different initial poster types: a user, and a page.
We discard ``celebrity'' users who may large followings like the most popular pages. 
Figure~\ref{fig:paper_shapes3}a shows that as the initial cascade structure becomes shallower, the proportion of cascades that double in size increases. To examine why this would be the case, we also examined the time needed for the 3rd reshare to happen (Figure~\ref{fig:paper_shapes3}c). For pages, shallower cascades tend to happen more rapidly, consistent with being initiated by a popular page and achieving a large number of reshares directly from its fans. Interestingly, the configuration having the second and third reshares stemming from the first reshare correspond to having a first resharer with many connections, and indicating that the initial poster is less popular, be it a page or user (Figure~\ref{fig:paper_shapes3}d).

Curiously, for user-started cascades, the star configuration tends to grow into the largest cascades, but is also the slowest. It also tends to correspond to the first resharer having a low degree, both for page and user roots. One might speculate that this pattern is indicative of the item's appeal to less well-connected users, who also happen to be more likely to reshare. In fact, a median resharer has 35 fewer friends than someone who is active on the site nearly every day. Thus, an item's appeal, rather than the initial network structure, may drive the eventual cascade size in the long run. 




\subsection{Predicting cascade structure}

The observations above naturally lead to the question of whether it is possible to predict future cascade structure.
In particular, we aim to distinguish cascades that spread like a virus in a shallow forest fire-like pattern (Figure \ref{fig:cascade_distances2}a) and cascades which spread in long, narrow string-like pattern (Figure \ref{fig:cascade_distances2}c).
As discussed earlier, this difference is related to the structural virality of a cascade and is quantified by the Wiener index.
Here, we observe $k=5$ reshares of a cascade and aim to predict whether the final cascade will have a Wiener index above or below the median. We obtain accuracy of 0.725 (F1 = 0.715, AUC = 0.796), while random guessing would, by construction, achieve accuracy of 0.5. 

\xhdr{Temporal and structural features are most predictive of structure}
For this task we expect structural features to be most important, while we expect temporal features not to be indicative of the cascade structure.
However, when we train the model on individual classes of features we surprisingly find that both temporal and structural features are almost equally useful in predicting cascade structure: 0.622 vs.\ 0.620.
Nevertheless, structural features remain individually more accurate ($\approx$ 0.58) 
and highly correlated ($0.161 \le |r| \le 0.255$) with the Wiener index.
Individually, one temporal feature, $\mathit{views}'_{1..k-1,k}$, is slightly more accurate (0.602) compared to the best-performing structural feature, $\mathit{outdeg}(\hat v_0)$ (0.600), but is significantly less correlated (0.041 vs.\ $-$0.255).
The two classes of features nicely complement each other, since when combined, accuracy increases to 0.72.

\xhdr{Cascade structure also becomes more predictable with increasing $k$}
Like for cascade growth prediction, our prediction performance improves as we observe more of the cascade, with accuracy linearly increasing from 0.724 when $k$ is 5 to 0.808 when $k$ is 100.
A linear relation also exists in the alternate task where we set the minimum cascade size $R$ to be 100, varying $k$ between 5 and 100.

\xhdr{Changes in feature importance}
As we increase $k$, we find that the structural features become highly correlated with the Wiener index, suggesting that the initial shape of a cascade is a good indicator of its final structure.
Rapidly growing cascades also result in final structures that are shallower---temporal features become more strongly correlated with the Wiener index as $k$ increases.
Unlike with cascade size, views were generally weakly correlated with structure, while content features had a weak, near-constant effect.
Nonetheless, some of these features still provided reasonable performance in the prediction task.

\xhdr{User vs.\ page-started cascades}
In predicting the shape of a cascade, we find that our overall prediction accuracy for pages is slightly higher (0.724) than for users (0.700).
While using only structural features alone results in a higher prediction accuracy for users (0.643) than for pages (0.601), user and content features are significantly more predictive of cascade structure in the case of pages.

To sum up, we find that predicting the shape of a cascade is not as hard as one might fear. Nevertheless, predicting cascade size is still much easier than predicting cascade shape, though classifiers for either achieve non-trivial performance.


\section{Predictability \& Content}
\label{sec:content}

\begin{figure}[t]
	\centering
	\includegraphics[width=\linewidth]{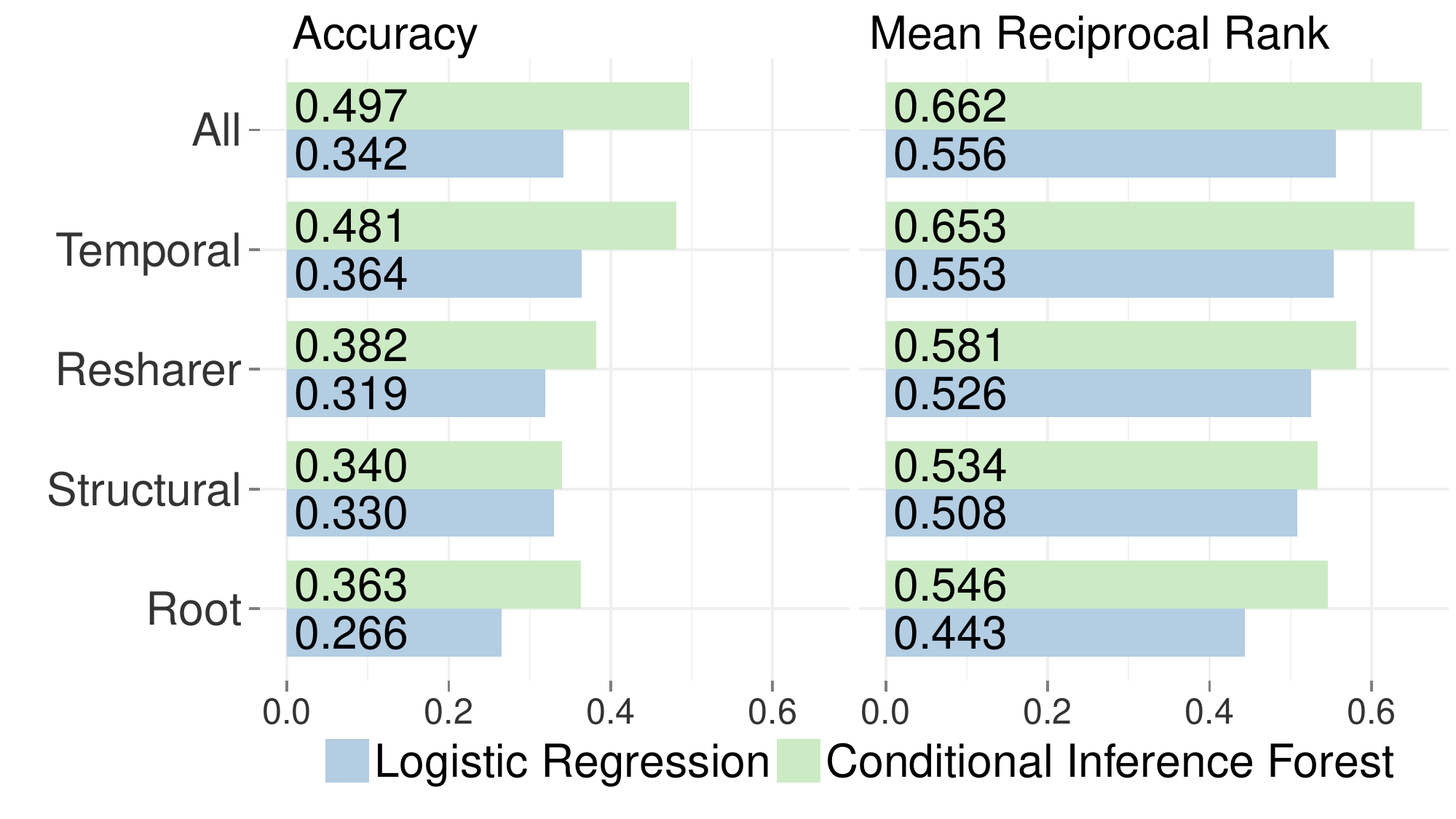}
	\caption{In predicting the largest cascade in clusters of 10 or more cascades of identical photos, we perform significantly above the baseline of 0.1.}
	\label{fig:cluster_prediction}
  \vspace{-0.5em}
\end{figure}

\subsection{Controlling for cascade content}\label{sec:content_control}
In our analyses thus far, we examined cascades of uploads of different photos, and tried to account for content differences by including photo and caption features. However, temporal and structural features may still capture some of the difference in content. 
Thus, we now study how well we can predict cascade size if we control for the content of the photo itself.
We consider {\em identical photos} uploaded to Facebook by different users and pages, which is not a rare occurrence.
We used an image matching algorithm to identify copies of the same image and place their corresponding cascades into clusters (983 clusters, $N_c =$ 38,073, $N_r =$ 12,755,621).
As one might expect, even the same photo uploaded at different times by different users can fare dramatically differently; a cluster typically consists of a few or even a single cascade with a large number of reshares, and many smaller cascades with few reshares.
The average Gini coefficient, a measure of inequality, is 0.787 ($\sigma =$ 0.104) within clusters.
Thus, a natural task is to try to predict the largest cascade within a cluster.
For every cluster we select 10 random cascades, placing the accuracy of random guessing at 10\%.

As shown in Figure \ref{fig:cluster_prediction}, in all cases we significantly outperform the baseline.
Using a random forest model, we can identify the most popular cascade nearly half the time (accuracy 0.497); a mean reciprocal rank of 0.662 indicates that this cascade also appears in the top two predicted cascades almost all the time.

In terms of feature importance we notice that best results are obtained using temporal features, followed by resharer, root node, and structural features.
Essentially, if one upload of the photo is initially spreading more rapidly than other uploads of the same photo, that cascade is also likely to grow to be the largest.
This points to the importance of landing in the right part of the network at the right time, as the same photo tends to have widely and predictably varying outcomes when uploaded multiple times.


\subsection{Feature importance in context}

\begin{figure}
\centering
\includegraphics[width=\linewidth]{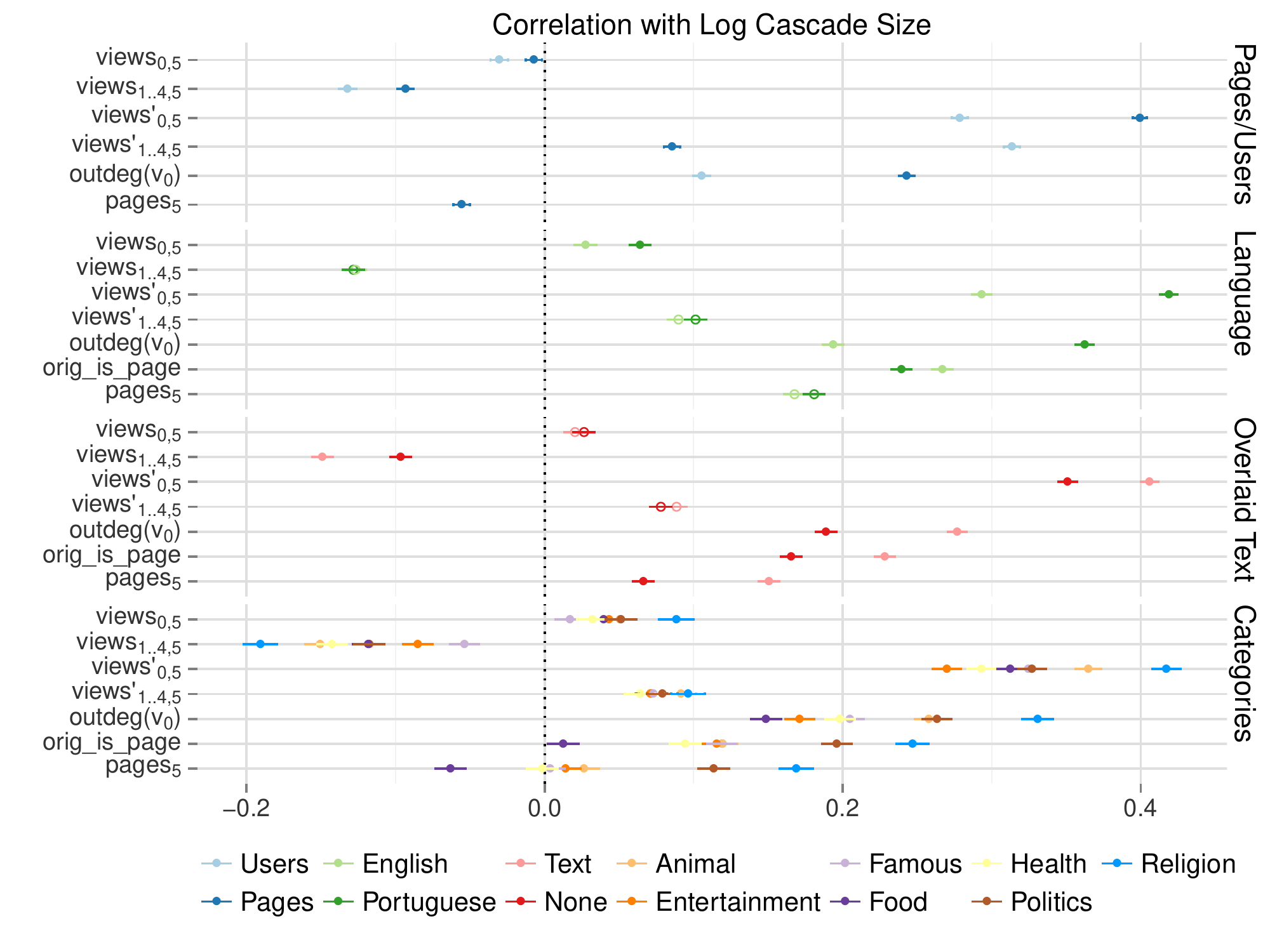}
\caption{The initial exposure of the uploaded photo and initial reshares serve to differentiate datasets from one another, as can be seen by comparing the correlation coefficients of each feature with the log cascade size. Solid circles indicate significance at $p <$ 10$^{\textrm{--3}}$, and lines through each circle indicate the 95\% confidence interval.}
\label{fig:cmp_page}
\vspace{-0.5em}
\end{figure}

Some features may be more or less important for our prediction tasks in different contexts. Figure~\ref{fig:cmp_page} shows how several features correlate with log-transformed cascade size when conditioned on one of four different variables, including
\begin{inparaenum}[(1)]
\item source node type---user vs page,
\item language---English versus Portuguese, the two most common languages of cascade root nodes in our dataset,
\item whether text is overlaid on a photo---a common feature of recent Internet memes, and
\item content category.
\end{inparaenum}
We determine content category by matching entities in photo captions to Wikipedia articles, and in turn articles to seven higher-level categories: animal, entertainment, politics, religion, famous people (excluding religious and political figures), food, and health.


Figure~\ref{fig:cmp_page} shows that the initial rate of exposure of the uploaded photo is generally more important for page cascades than for user cascades ($\mathit{views}'_{0,5}$).
This is likely due to the higher variance in the distribution of the number of followers for a user versus a page. For page cascades in our sample, the median number of followers is 73,855 with a standard deviation of 675,203, while for users at the root of cascades the median number of friends and subscribers is 1,042 with a standard deviation of 26,482. Though rate of exposure to the original photo is more important for pages, we see that rate of exposure to the initial reshares ($\mathit{views}'_{1..4,5}$) is much more important for user cascades.

The number and rate of views also act to differentiate topical categories, with religion having the highest correlation between views and cascade size.
Correlation for the rate of views of the uploaded photo is also higher for those with a Portuguese-speaking root node as opposed to an English one.
The feature $\mathit{outdeg}(v_0)$ indicates the ability of the root to broadcast content, and we see this playing an important role for page cascades, Portuguese content, photos with text, and religious photos. This indicates that much of the success of these cascades is related to the root nodes being directly connected to large audiences.


In addition to the analysis of Figure~\ref{fig:cmp_page}, we also examined how the features correlate with the structural virality of the final cascades. (Each of the reported correlation coefficient comparisons that follow are significant at $p <$ 10$^{\textrm{--3}}$ using a Fisher transformation.) Photos relating to food differ significantly from all other categories in that features of the root, such as $\mathit{outdeg}(v_0)$, are less negatively correlated ($>$--0.18 vs. --0.11), and depth features, such as $\mathit{depth}_k^\mathit{avg}$, are less positively correlated ($>$0.18 vs. 0.11). This relationship also holds for English compared to Portuguese photos. While users with many friends or followers are more likely to generate cascades of larger size and greater structural virality, pages with many fans create cascades of larger size, although not necessarily greater virality (0.05 vs. --0.01).
However, if the initial structure of a cascade is already deep, the final structure of the cascade is likely to have greater structural virality for both user and page-started cascades ($>$0.16).
A user-started cascade whose initial reshares are viewed more quickly is also more likely to become viral than that for a page-started cascade (0.23 vs. 0.06).

\section{Discussion \& Conclusion}
\label{sec:conclusion}
This paper examines the problem of predicting the growth of cascades over social networks. Although predictive tasks of similar spirit have been considered in the past, 
we contribute a novel formulation of the problem which does not suffer from skew biases. Our formulation allows us to study predictability throughout the life of a cascade. We examine not only how the predictability changes as more and more of the cascade is observed (it improves), but also how predictable large cascades are if we only observe them initially (larger cascades are more difficult to predict). While some features, e.g., the average connection count of the first $k$ resharers, have increasing predictive ability with increasing $k$, others weaken in importance, e.g., the connectivity of the root node. We find that the importance of features depends on properties of the original upload as well: the topics present in the caption, the language of the root node, as well as the content of the photo.

Despite the rich set of results we were able to obtain, there are some limitations to this study. 
Most importantly, the study was conducted entirely with Facebook data and only with photos.
Still, one advantage of this is the scale of the medium; hundreds of millions of photos are uploaded to Facebook every day, and photos, more than other content types, tend to dominate reshares.
This also gives us high-fidelity traces of how the photo moves within Facebook's ecosystem, which allows us to precisely overlay the spreading cascade over the social network. Moreover, we are able to identify uploads of the same photo and track them individually. This eliminates the concern of shares being driven by an external entity and only appearing to be spreading over the network. 
Instead, external drivers benefit our study by creating independent `experiments' where the same photo gets multiple chances to spread, helping us control for the role of content in some of our experiments.
Another disadvantage of our setup is that diffusion within Facebook is driven by the mechanics of the site. The distinction between pages and users is specific to Facebook, as are the mechansisms by which users interact with content, e.g., liking and resharing. Despite these limitations, we believe the results give general insights which will be useful in other settings. 

\begin{figure}[t]
\centering
\includegraphics[width=\linewidth]{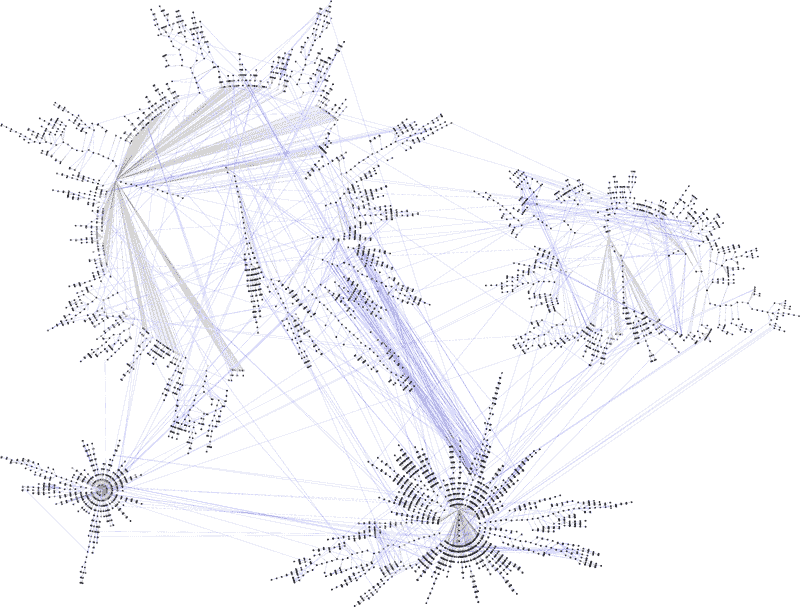}
\caption{There is considerable overlap in friendship edges (blue) between four independent cascades of the same photo.}
\label{fig:cascade_overlap}
\vspace{-0.5em}
\end{figure}

The present work only examines each cascade independently from others.
Future work should examine interactions between cascades, both between different content competing for the same attention, and between the same content surfacing at different times and in different parts of the network.
We found that when the same photo is uploaded at least 10 times, the largest cascade was twice as likely to be among the first 20\% of uploads than the last 20\%. 
Similarly, for photos uploaded 20 times, the largest cascade was 2.3 times as likely to be among the first 20\% than the last.
Figure~\ref{fig:cascade_overlap} shows the friendship edges between users participating in different cascades of a single, specific photo. The high connectivity {\em between} different cascades demonstrates that users are likely being exposed to the same photo via different cascades, which could be a contributing factor in why earlier uploads of the same photo tend to generate larger cascade than later ones. Between-cascade dynamics like this should provide ample opportunities for further research.

Addressing questions like these will lead to a richer understanding of how information spreads online and pave the way towards better management of socially shared content and applications that can identify trending content in its early stages.



\newpage
\bibliographystyle{abbrv}
\bibliography{refs}


\end{document}